\documentclass[12pt,letter]{article}
\usepackage{amsfonts,amssymb,amsmath,epsfig}

\numberwithin{equation}{section}
\begin{document}
\newcommand{\tr}{\operatorname{tr}}
\newcommand{\ds}{\displaystyle}
\newcommand{\scr}{\scriptstyle}
\newcommand{\sscr}{\scriptstyle}
\newcommand{\e}{\operatorname{e}}
\newcommand{\mustbe}{\stackrel{!}{=}}
\newcommand{\rt}{(r,\vartheta)}
\newcommand{\rts}{(r',\vartheta')}

\newcommand{\sdot}{\,{\scriptscriptstyle{}^{\bullet}}\,}

\newcommand{\trip}{{}^3S_1}
\newcommand{\sing}{{}^1S_0}
\newcommand{\A}{\mathcal{A}}

\newcommand{\ap}{{}^{(a)}\varphi_{\pm}(\vec{r})}
\newcommand{\app}{{}^{(a)}\varphi_{+}(\vec{r})}
\newcommand{\apm}{{}^{(a)}\varphi_{-}(\vec{r})}

\newcommand{\sapm}{{}^{(a)} \varphi_{-}^{\dagger}(\vec{r})}
\newcommand{\sapp}{{}^{(a)} \varphi_{+}^{\dagger}(\vec{r})}

\newcommand{\ipp}{{}^{(1)}\varphi_{+}(\vec{r})}
\newcommand{\ipm}{{}^{(1)}\varphi_{-}(\vec{r})}
\newcommand{\iipp}{{}^{(2)}\varphi_{+}(\vec{r})}
\newcommand{\iipm}{{}^{(2)}\varphi_{-}(\vec{r})}
\newcommand{\sPh}{{\stackrel{*}{\Phi}}(\vec{r}_1,\vec{r}_2)}
\newcommand{\sph}{{\stackrel{*}{\varphi}}}
\newcommand{\sphi}{{\stackrel{*}{\varphi}_1}(\vec{r}_1)}
\newcommand{\sphii}{{\stackrel{*}{\varphi}_2}(\vec{r}_2)}
\newcommand{\sphis}{{\stackrel{*}{\varphi}_1}(\vec{r}\,')}
\newcommand{\sphiis}{{\stackrel{*}{\varphi}_2}(\vec{r}\,')}

\newcommand{\aRpm}{{}^{(a)\!}\tilde{R}_\pm}
\newcommand{\aRp} {{}^{(a)\!}\tilde{R}_+{}}
\newcommand{\aRm}{{}^{(a)\!}\tilde{R}_-{}}
\newcommand{\iRp}{{}^{(1)\!}\tilde{R}_+{}}
\newcommand{\iRm}{{}^{(1)\!}\tilde{R}_-{}}
\newcommand{\iiRp}{{}^{(2)\!}\tilde{R}_+{}}
\newcommand{\iiRm}{{}^{(2)\!}\tilde{R}_-{}}
\newcommand{\ntR}{{}^{\{0\}}\!\tilde{R}}
\newcommand{\aSpm}{{}^{(a)\!}\tilde{S}_\pm}
\newcommand{\aSp}{{}^{(a)\!}\tilde{S}_+{}}
\newcommand{\aSm}{{}^{(a)\!}\tilde{S}_-{}}
\newcommand{\iSp}{{}^{(1)\!}\tilde{S}_+{}}
\newcommand{\iSm}{{}^{(1)\!}\tilde{S}_-{}}
\newcommand{\iiSp}{{}^{(2)\!}\tilde{S}_+{}}
\newcommand{\iiSm}{{}^{(2)\!}\tilde{S}_-{}}

\newcommand{\ikn}{{}^{(1)}\!k_0(\vec{r})}
\newcommand{\ikp}{{}^{(1)}\!k_\phi(\vec{r})}
\newcommand{\iikp}{{}^{(2)}\!k_\phi(\vec{r})}

\newcommand{\iikn}{{}^{(2)}\!k_0(\vec{r})}
\newcommand{\ikns}{{}^{(1)}\!k_0(\vec{r}\,')}
\newcommand{\iikns}{{}^{(2)}\!k_0(\vec{r}\,')}

\newcommand{\itkn}{{}^{(1)}\!\tilde{k}_0}
\newcommand{\iitkn}{{}^{(2)}\!\tilde{k}_0}

\newcommand{\itkp}{{}^{(1)}\!\tilde{k}_\phi}
\newcommand{\iitkp}{{}^{(2)}\!\tilde{k}_\phi}

\newcommand{\ajn}{{}^{(a)}\!j_0(\vec{r})}
\newcommand{\akn}{{}^{(a)}\!k_0\left(\vec{r}\right)}
\newcommand{\sakn}{{}^{(a)}\!k_0\left(\vec{r}\,'\right)}
\newcommand{\btkn}{{}^{(b)}\!\tilde{k}_0\left(r\right)}
\newcommand{\bkn}{{}^{(b)}\!k_0\left(\vec{r}\right)}
\newcommand{\bknn}{{}^{(b)}\!k_0\left(r\right)}
\newcommand{\btknn}{{}^{(b)}\!\tilde{k}_0\left(r\right)}
\newcommand{\aki}{{}^{(a)}\!k_1(\vec{r})}
\newcommand{\akp}{{}^{(a)}\!k_\phi(\vec{r})}
\newcommand{\akv}{\vec{k}_a(\vec{r})}

\newcommand{\aAr}{{}^{(a)}\!A_0}
\newcommand{\aAe}{{}^{[a]}\!A_0}
\newcommand{\aAn}{{}^{(a)}\!A_0(\vec{r})}
\newcommand{\pAn}{{}^{(\textrm{p})}\!A_0(\vec{r})}
\newcommand{\ppAn}{{}^{[\textrm{p}]}\!A_0}
\newcommand{\nF}{{}^{\{0\}}\!F}
\newcommand{\nAn}{{}^{\{0\}}\!A_0}
\newcommand{\ppF}{{}^{[\textrm{p}]}\!F(r)}
\newcommand{\ppFn}{{}^{[\textrm{p}]}\!F(0)}
\newcommand{\iAn}{{}^{(1)}\!\!A_0(\vec{r})}
\newcommand{\iiAn}{{}^{(2)}\!\! A_0 (\vec{r}) }
\newcommand{\fAn}{{}^{(1/2)}\!\! A_0 (\vec{r}) }

\newcommand{\aVHS}{{}^{(a)}V_{\textrm{HS}}(\vec{r})}
\newcommand{\iVHS}{{}^{(1)}V_{\textrm{HS}}(\vec{r})}
\newcommand{\iiVHS}{{}^{(2)}V_{\textrm{HS}}(\vec{r})}

\newcommand{\onp}{\omega_0^\textrm{(+)}}
\newcommand{\onm}{\omega_0^\textrm{(-)}}
\newcommand{\oip}{\omega_1^\textrm{(+)}}
\newcommand{\oim}{\omega_1^\textrm{(-)}}
\newcommand{\onpm}{\omega_0^{(\pm)}}
\newcommand{\oipm}{\omega_1^{(\pm)}}

\renewcommand{\L}{\mathcal{L}}
\newcommand{\B}{\mathcal{B}}
\newcommand{\as}{\alpha_{\textrm{s}}\,}
\newcommand{\aB}{a_{\textrm{B}}\,}
\newcommand{\ab}{a_{\textrm{b}}\,}
\newcommand{\Bstar}{\overset{*}{B}}
\newcommand{\Gstar}{\overset{*}{G}}
\newcommand{\gstar}{\overset{*}{g}}
\newcommand{\CC}{\mathbb{C}}
\newcommand{\GG}{\textnormal I \! \Gamma}
\newcommand{\D}{\mathcal{D}}
\newcommand{\F}{\mathcal{F}}
\renewcommand{\H}{\mathcal{H}}
\newcommand{\I}{\mathcal{I}}
\newcommand{\J}{\mathcal{J}}
\newcommand{\LRST}{\mathcal{L}_{\textrm{RST}}}
\newcommand{\LD}{\mathcal{L}_{\textrm{D}}}
\newcommand{\LG}{\mathcal{L}_{\textrm{G}}}
\newcommand{\U}{\mathcal{U}}
\newcommand{\M}{\mathcal{M}}
\newcommand{\T}{\mathcal{T}}
\newcommand{\Mpe}{M_\textrm{p/e}\,}
\newcommand{\Mp}{M_\textrm{p}\,}
\newcommand{\Mee}{M^\textrm{(e)}}
\newcommand{\Me}{M_\textrm{e}\,}
\newcommand{\MT}{M_\textrm{T}}
\newcommand{\tMT}{\tilde{M}_\textrm{T}}
\newcommand{\Mei}{M_\textrm{I}^\textrm{(e)}}
\newcommand{\Meii}{M_\textrm{II}^\textrm{(e)}}
\newcommand{\Meiii}{M_\textrm{I/II}^\textrm{(e)}}
\newcommand{\Mmi}{M_\textrm{I}^\textrm{(m)}}
\newcommand{\Mmii}{M_\textrm{II}^\textrm{(m)}}
\newcommand{\Mmiii}{M_\textrm{I/II}^\textrm{(m)}}
\newcommand{\TT}{{}^{(\textrm{T})}T}
\newcommand{\DT}{{}^{(\textrm{D})}T}
\newcommand{\GT}{{}^{(\textrm{G})}T}
\newcommand{\hER}{\hat{E}_\textrm{R}}
\newcommand{\heER}{\hat{E}_\textrm{R}^\textrm{(e)}}
\newcommand{\hmER}{\hat{E}_\textrm{R}^\textrm{(m)}}

\newcommand{\ED}{{E_{\textrm{D}}}}
\newcommand{\EG}{E_{\textrm{G}}}
\newcommand{\Ec}{E_{\textrm{conv}}}
\newcommand{\cEc}{{\cal E}_{\textrm{conv}}}
\newcommand{\ET}{E_{\textrm{T}}}
\newcommand{\Epot}{ {\cal E}_{\textrm{pot}} }
\newcommand{\tET}{\tilde{E}_\textrm{T}}
\newcommand{\tETi}{\tilde{E}_\textrm{T}^{(|)}}
\newcommand{\tETii}{\tilde{E}_\textrm{T}^{(||)}}
\newcommand{\tEnT}{ {\tilde{E}^{(0)}_\textrm{T}} }
\newcommand{\tEnTmin}{ {\tilde{E}^{(0)}_\textrm{T,min}} }
\newcommand{\ETiii}{E_\textrm{T}^{(|||)}}
\newcommand{\Ekin}{{E_\textrm{kin}}}
\newcommand{\cEk}{{\cal{E}_\textrm{kin}}}
\newcommand{\ES}{{E_{\textrm{S}}}}
\newcommand{\EW}{{E_\textrm{W}}}
\newcommand{\HS}{H_{\textrm{S}}}
\newcommand{\hHin}{\hat{H}_{\textrm{in}}}
\newcommand{\hHS}{\hat{H}_{\textrm{S}}}
\newcommand{\WRST}{W_{\textrm{RST}}}
\newcommand{\WS}{W_{\textrm{S}}}
\newcommand{\lG}{{\lambda_\textrm{G}}}
\newcommand{\lGe}{ {\lambda_\textrm{G}^{(\textrm{e})}}\!}
\newcommand{\lGm}{  {\lambda_\textrm{G}^{(\textrm{m})}}\!\!}
\newcommand{\lGma}{  {\lambda_{\textrm{G}(a)}^{(\textrm{m})}}\!\!}
\newcommand{\lGem}{ {\lambda_\textrm{G}^{(\textrm{e/m})}}}
\newcommand{\lGema}{ {\lambda_{\textrm{G}(a)}^{(\textrm{e/m})}}}
\newcommand{\lD}{{\lambda_\textrm{D}}}
\newcommand{\lS}{{\lambda_\textrm{S}}}
\newcommand{\np}{{n_\textrm{p}}}
\newcommand{\NN}{\mathbb{N}}
\newcommand{\ND}{{N_\textrm{D}}}
\newcommand{\NDn}{{N_\textrm{D}^\textrm{(0)}}\!\!}
\newcommand{\NGn}{{N_\textrm{G}^\textrm{(0)}}\!\!}
\newcommand{\NGe}{{N_\textrm{G}^\textrm{(e)}}\!\!}
\newcommand{\NGm}{{N_\textrm{G}^\textrm{(m)}}\!\!\!}
\newcommand{\LDk}{{\LD^{(\mathrm{kin})}}}
\newcommand{\LDe}{{\LD^{(\mathrm{e})}}}
\newcommand{\LDm}{{\LD^{(\mathrm{m})}}}
\newcommand{\LDM}{{\LD^{(\mathrm{M})}}}
\newcommand{\Tkin}{{T_\textrm{kin}}}
\newcommand{\oWD}{{\overset{\circ}{W}\!}_\textrm{D}}
\newcommand{\oWG}{{\overset{\circ}{W}\!}_\textrm{G}}
\newcommand{\oWRST}{{\overset{\circ}{W}\!}_\textrm{RST}}
\newcommand{\oWRSTe}{{\overset{\circ}{W}}{}^{\textrm{(e)}}_\textrm{RST}}
\renewcommand{\S}{\mathcal{S}}
\newcommand{\Ph}{\Phi(\vec{r}_1,\vec{r}_2)}
\newcommand{\VN}{V_\NN}
\newcommand{\TN}{T_\NN}
\newcommand{\cTN}{\mathcal{T}_\NN}
\newcommand{\PN}{P_\NN}
\newcommand{\pN}{(p_0,p_1,\ldots p_\NN) }
\newcommand{\cVN}{\mathcal{V}_\NN}
\newcommand{\Z}{\mathcal{Z}}

\title{\bf Non-Relativistic Positronium Spectrum\\ in\\ Relativistic Schr\"odinger Theory}
\author{M.\ Mattes and M.\ Sorg\\[1cm] II.\ Institut f\"ur Theoretische Physik der
Universit\"at Stuttgart\\ Pfaffenwaldring 57 \\ D 70550 Stuttgart, Germany\\Email: sorg@theo2.physik.uni-stuttgart.de\\http://www.theo2.physik.uni-stuttgart.de/institut/sorg/publika.html\vspace{-2mm}}
\date{ }
\maketitle
\begin{abstract}
  The lowest energy levels of positronium are studied in the non-relativistic
  approximation within the framework of Relativistic Schr\"o\-dinger Theory (RST). Since
  it is very difficult to find the exact solutions of the RST field equations (even in the
  non-relativistic limit), an approximation scheme is set up on the basis of the
  hydrogen-like wave functions (i.e.\ polynomial times exponential). For any approximation
  order~$\NN\ (\NN=0,1,2,3,\ldots)$ there arises a spectrum of approximate RST solutions
  with the associated energies, quite similarly to the conventional treatment of
  positronium in the standard quantum theory (Appendix). For the lowest approximation
  order~$(\NN=0)$ the RST prediction for the \emph{groundstate} energy exactly agrees with the
  conventional prediction of the standard theory. However for the higher approximation
  orders~$(\NN=1,2,3)$, the corresponding RST prediction differs from the conventional
  result by (roughly)~$0,9\ [eV]$ which confirms the previous estimate of the error being
  due to the use of the spherically symmetric approximation. The excited states require
  the application of higher-order approximations~$(\NN>>3)$ and are therefore not
  adequately described by the present orders~$(\NN\le 3)$.
\vspace{4cm}
\noindent

\end{abstract}


\section{Introduction and Survey of Results}
\indent

Apparently the general belief says that the standard quantum theory is unique in the sense
that no other theory can quantitatively predict the phenomena of the microscopic world
more accurately than just that conventional theory. But this does logically not exclude
the possibility, that other forms of quantum theory may perhaps exist which have the same
predictive power: \emph{``The natural meaning of the claim that quantum theory provides
  for a complete scientific account of atomic phenomena is that no theoretical
  construction can yield experimentally verifiable predictions about atomic phenomena that
  cannot be extracted from a quantum theoretical description''}~\cite{St}. It is true,
this claim is surely intended to refer to the standard form of quantum theory; however, if
one could find an alternative formalism with the same predictive potentiality as the
standard theory, this would be nevertheless of considerable relevance because it could
shed new light on the old interpretation problems of the standard theory.

Perhaps the recently established Relativistic Schr\"odinger Theory (RST) does represent
such an alternative candidate~\cite{BeSo,MaSo}. This theory is not of probabilistic but
rather of fluid-dynamic character and therefore resembles rather the well-known density
functional theory~\cite{PY,DrGr} than the usual statistical approach. Naturally, such a
philosophical difference must then entail also certain consequences for the mathematical
structure of both theoretical approaches: the~$N$-particle systems are described in RST
by the Whitney sum of the one-particle bundles, not by the tensor product of the
one-particle Hilbert spaces as in the standard theory. However it is well-known that also
such a fluid-dynamic approach as the density functional theory can predict atomic and
molecular data with the same accuracy as the standard probabilistic approach, and this
should justify the endeavors for further elaborating those fluid-dynamic theories.

For instance, it has recently been observed that RST, when being applied to the
non-relativistic positronium problem, predicts \emph{exactly} the same groundstate energy
as the conventional quantum mechanics~\cite{MaSo}. Recall here that this conventional
prediction is based upon a very simple argument: Assuming (for the non-relativistic
situation) that the electron and positron undergo the electromagnetic interaction via the
usual Coulomb potential, the corresponding two-particle Hamiltonian~$\hat{H}$ reads
\begin{equation}
  \label{eq:I.1}
  \hat{H}=\frac{\vec{p}^{\;2}_1}{2M} + \frac{\vec{p}^{\;2}_2}{2M} -
  \frac{e^2}{||\vec{r}_1-\vec{r}_2||}\ .
\end{equation}
This Hamiltonian can easily be transformed to the sum of an external part (due to the
center-of-mass motion) and an internal part~$\hHin$~\cite{Me}
\begin{equation}
  \label{eq:I.2}
  \hHin = \frac{\vec{p}^{\;2}}{2m} - \frac{e^2}{r}
\end{equation}
which then yields the well-known conventional spectrum of the internal excitations:
\begin{equation}
  \label{eq:I.3}
  \Ec = -\frac{e^2}{4\aB}\cdot\frac{1}{(\np+1)^2}\simeq-\frac{6,8029\ldots}{(\np+1)^2}\ \ [eV]\ ,
\end{equation}
with the principal quantum number~$n_p$ being some integer~$(\np=0,1,2,3\ldots)$ and~$\aB$
the Bohr radius
\begin{equation}
  \label{eq:I.4}
  \aB = \frac{\hbar^2}{Me^2}\ .
\end{equation}
Indeed, the eigenvalue equation of the internal Hamiltonian (\ref{eq:I.2}) looks very
simple
\begin{gather}
  \label{eq:I.5}
  \hHin\phi\equiv -\left(\frac{\hbar^2}{2m}\Delta+\frac{e^2}{r} \right)\phi =
  \Ec\phi\\*
  \big(m=\frac{M}{2}\ldots\text{reduced mass} \big)\ ,\notag
\end{gather}
and can easily be solved by the product of a polynomial times an exponential (cf.~Appendix).

But the problem for RST consists now just in that \emph{exact} coincidence of its
groundstate prediction~\cite{MaSo,BMS} with the conventional result (\ref{eq:I.3})
for~$n_p=0$!  Namely, whereas that conventional result owns the status of absolute
exactness within the standard framework (apart from relativistic and QED
corrections~\cite{Ka}), its RST counterpart is only of \emph{approximative} character,
even within the non-relativistic approximation of RST. The point here is that even the
non-relativistic approximation of RST does not lead to a spherically-symmetric potential
(such as the conventional Coulomb potential occurring in the internal Hamiltonian
(\ref{eq:I.5})); but rather the non-relativistic RST potential must be further truncated
to the spherically-symmetric form in order that the corresponding groundstate prediction
can exactly agree with the conventional result (\ref{eq:I.3}) for~$n_p=0$. But even if
such a truncation to the spherical symmetry of the RST groundstate configuration is
accepted, the corresponding eigenvalue problem cannot be solved exactly; but rather one
has to resort to some approximative procedure whose first step then yields the mentioned
coincidence of the RST and conventional predictions, see refs.~\cite{MaSo,BMS}.

This circumstance, i.e.\ the numerical coincidence of the \emph{``exact''} conventional
result and the \emph{approximative} RST result in the non-relativistic domain, is just the
reason why one cannot be completely satisfied with the present situation. Namely, if one
wishes to stick to the claim that the \emph{true} nonrelativistic limit of RST should
agree to a high precision with the conventional predictions, then the spherically
symmetric RST approximation of that true limit should differ by a certain finite amount
from both the conventional results and the true RST prediction. This difference, being due
to the \emph{spherically symmetric approximation} of the non-relativistic interaction
potential, has been estimated as (roughly)~$1\ [eV]$, see the appendix of
ref.~\cite{BMS}. Therefore one has to expect that the spherically symmetric
(non-relativistic) RST limit of the positronium groundstate should differ from the
conventional prediction (\ref{eq:I.3}) by roughly~$1\ [eV]$. The present paper aims at the
verification of this energy difference and thus upholds the possibility that the true
(i.e.\ anisotropic) non-relativistic RST limit coincides practically with the conventional
predictions. (The common difference of both theoretical approaches relative to the
experimental data would then refer exclusively to the relativistic corrections).

In this sense, the main result of the present paper consists in the elaboration of the
expected energy difference of roughly~$1\ [eV]$. This goal is attained by setting up a
rigorous approximation formalism so that the incidental numerical coincidence of the RST
and conventional predictions do appear only in the zero-order approximation. In the
first-order approximation~$(\NN=1)$, there emerges an energy difference of (roughly)~$0,9\
[eV]$ which thus verifies the expectations. Since this energy difference persists also in
the higher-order approximations~$(\NN=2,3)$ it seems justified to hypothesize that the true
non-relativistic RST limit (which includes also the anisotropic effects) will actually
meet with the corresponding non-relativistic predictions of the conventional theory.

These results are worked out in the following arrangement:

First, a brief sketch of the general relativistic two-particle theory is presented in
\textbf{Sect.~II}. Here the principal RST dynamics is displayed, including the
specialization to the stationary bound states. As a preparation of the subsequent
treatment of the positronium groundstate, the relativistic field equations are then
transcribed to their non-relativistic and spherically symmetric approximations.

Next, the RST \emph{principle of minimal energy}~\cite{MaSo} is presented in
Sect.~\textbf{III}. This variational principle is the crucial point for the subsequent
calculation of the positronium groundstate energy since the exact solutions of the RST
field equations are very hard to obtain, so that one has to be satisfied with such an
approximate variational technique. But fortunately this is most adequately provided by
that principle of minimal energy whose variational equations consist in the coupled matter
and gauge field equations (see the Dirac equations (\ref{eq:II.45a})-(\ref{eq:II.45b}) for
the matter subsystem and the Poisson equations (\ref{eq:II.43a})-(\ref{eq:II.43d}) for the
gauge field subsystem). Both subsystem variations do need a constraint, i.e.\ the wave
function normalization for the matter fields (\ref{eq:III.11}) and the \emph{Poisson
  identities} (\ref{eq:III.19}) for the gauge fields. The latter identities refer to the
numerical equality of the pure gauge field energy~$\heER$ and the interaction energy~$\Mee
c^2$ of the matter and gauge fields (\emph{mass equivalent}). The non-relativistic version
of the matter part of this variational principle turns out as nothing else than the
well-known Ritz principle of conventional quantum theory. The Poisson identities do not
only ensure the deduction of the gauge field equations from the principle of minimal
energy, but provide also a nice consistency check for the concrete calculation of the
electric gauge field energy~$\heER$. Namely, the latter energy essentially consists of
three parts: the groundstate contribution, the quadratic terms and the quartic terms, see
equation (\ref{eq:III.53}). Since the same subdivision does hold also for the electric
\emph{mass equivalent}~$\Mee c^2$ (\ref{eq:III.61}), the Poisson identity implies the
separate numerical equality of all three contributions, see equations
(\ref{eq:III.62})-(\ref{eq:III.67}) . Clearly such a set of integral identities is a
rigorous consistency test for the correct computation of the gauge field energy with
respect to the chosen trial functions.

With all these preparations, the proper goal of the paper can be tackled in
\textbf{Sect.~IV}: Selecting as trial functions the hydrogen-like wave
functions~$\tilde{R}(r)$ (see equation (\ref{eq:III.35}) together with (\ref{eq:IV.1})),
one can study the action of the non-relativistic energy functional~$\tEnT$
(\ref{eq:III.32}) on this set of functions, of course in order to determine the stationary
points and especially the minimally possible energy value (groundstate). Since the
hydrogen-like wave functions are the product of an exponential times a polynomial of
degree~$\NN$, these trial functions may be classified just by the degree~$\NN$. In this
way there arises an energy spectrum for any order~$\NN\ (\NN=0,1,2,3,\ldots)$ each of
which has a finite number of members. The lowest-energy state is of course the groundstate
which is our main concern because its RST energy has been found to agree exactly with the
conventional prediction (\ref{eq:I.3}). In order to better understand this strange
agreement of the RST and conventional groundstate predictions, the same variational
technique is applied in the appendix also to the conventional positronium problem
(\ref{eq:I.5}); and the comparison of both approaches yields the following results:

The lowest-order spectrum~$(\NN=0)$ consists exclusively of the groundstate whose energy
is found in this order to be the same for both approaches, cf.~the RST result
(\ref{eq:IV.36}) and the conventional result (\ref{eq:A.18}). For the higher-order
spectra~$(\NN\ge 1)$ it is important to observe that the constraint of wave function
normalization forces one to look for the stationary points of the energy function on an
$\NN$-dimensional sphere~$S^{\NN}$. This is not quite trivial from the topological point
of view since the well-known Morse inequalities~\cite{Mi,Mo} imply certain restrictions
upon the number of stationary points which a continuously differentiable function can own
on a given compact manifold. For instance, on an $\NN$-dimensional
torus~$T^{\NN}=S^1\times S^1\times\ldots S^1$ the minimally possible number of stationary
points is~$2^{\NN}$~\cite{NaSe}. But fortunately the proposition of Reeb~\cite{CDD} ensures for
the spheres~$S^\NN$ that the minimally possible number is 2; and therefore the
conventional approach for~$\NN=1$ ($\leadsto$~Appendix) can produce two stationary points
(i.e.\ minimum and maximum) of the energy function on the circle~$S^1$, see
fig.~A.1. Clearly this corresponds to the groundstate and first excited state as
\emph{exact} solutions of the internal Schr\"odinger equation (\ref{eq:I.5}). Furthermore,
the Reeb proposition admits that the general conventional energy function (\ref{eq:A.14})
has \emph{exactly}~$\NN+1$ stationary points on~$S^\NN$ which generate the conventional spectrum
(\ref{eq:I.3}) with~$0\le n_p\le \NN$; see the exemplification for~$\NN=2$ in the Appendix.

A very pleasant feature of the conventional spectrum refers to the fact, that all
stationary points of~$\Ec\{\NN\}$ (\ref{eq:A.14}) on~$S^\NN$ are present also
on~$S^{\NN+1}$, i.e.\ the property of stationarity is not lost by the embedding~$S^\NN\to
S^{\NN+1}$! This is in contrast to the situation with RST. For instance, the RST
groundstate energy (\ref{eq:IV.36}) for~$\NN=0$ actually emerges also in the next
order~$\NN=1$ in form of the field configuration with energy~$E_\mathrm{1a}$
(\ref{eq:IV.45a}), but this corresponds in this order~$\NN=1$ to a maximum (or saddle
point, resp.) and not to a minimum as in the lowest order~$\NN=0$. Therefore the agreement
of the conventional result with the RST groundstate energy (\ref{eq:IV.36}) and
(\ref{eq:IV.45a}) is not ``stable'' within the RST approach, as is the case within the
conventional approach. The concept of ``stability'' refers here to the property of a
stationary point to be preserved by the embedding~$S^\NN \to S^{\NN+1}$. The instability
of the ``exact'' RST energy (\ref{eq:IV.36}) and (\ref{eq:IV.45a}) for the positronium
groundstate is revealed by the next higher embedding~$S^1\to S^2$. Indeed, this embedding
does preserve three of the four stationary points of order~$\NN=1$, but the ``exact''
configuration with~$E_\mathrm{1a}$ (\ref{eq:IV.45a}) is lost and there arise six new
stationary points with different energies~$E_\mathrm{2e}-E_\mathrm{2j}$
(\ref{eq:IV.57a})-(\ref{eq:IV.57f}).

However concerning the groundstate, this situation just meets with the expectations
concerning the numerical accuracy of the present spherically symmetric
approximation. Since it has been estimated that this kind of approximation generates a
deviation of (roughly)~$1\ [eV]$ from the true (but non-relativistic) RST predictions for
the positronium groundstate~\cite{BMS}, it is necessary that the present RST prediction
deviates from the conventional result (\ref{eq:A.18}) by roughly~$1\ [eV]$. Now the
minimal value of the RST energy function~$\tEnT$ (\ref{eq:IV.24}) is found here in the
order~$\NN=1$ as~$E_\mathrm{1d}$ (\ref{eq:IV.46b}), and this survives also the passage to
the next order~$\NN=2$. It is true, the case with the ``exact'' RST configuration~(1a)
(\ref{eq:IV.45a}) demonstrates that a stationary point of a lower order may survive to
some higher order but nevertheless vanishes during the passage to the subsequent orders,
but the relatively small change of the RST groundstate energy for~$\NN=0\ (\leadsto
E_0=-6,8029\ldots[eV])$, cf.~(\ref{eq:IV.36}), to its value for~$\NN=1$
and~$\NN=2$~$(\leadsto E_\mathrm{1d}=-7,675\ldots[eV])$ suggests to hypothesize that the
``true'' (i.e.~$\NN\to\infty$) RST prediction (in the spherically-symmetric approximation)
actually remains sufficiently close to the exact conventional value (\ref{eq:IV.36});
i.e.\ more precisely: the isotropic RST prediction remains within the limits which are put
by the anisotropic estimate~\cite{BMS} of~$1\ [eV]$, so that the ``true'' RST prediction
(i.e.\ with observation of the anisotropy) practically can agree with the conventional
result! Clearly in order to verify this rigorously, one would have to go explicitly
through the higher orders up to~$\NN\to\infty$. Here one could then also study the more
mathematical questions of how many and which of the~$\NN$-order stationary points do
survive the passage to the next order~$(\NN+1)$, thus arriving for~$\NN\to\infty$ at the
exact RST spectrum in the spherically-symmetric approximation. The RST spectrum of
order~$\NN=3$ is sketched in fig.~2 (at the end of the main text).


\section{Relativistic Two-Particle Theory}
\indent

Although the subsequent discussion refers exclusively to the \emph{non-relativistic} limit
of RST, it is important to realize that this theory is of \emph{truly relativistic}
character. Therefore the non-relativistic results displayed below are to be understood in
the sense that they represent merely the non-relativistic limit forms of the corresponding
relativistic outcomes which will however not be presented in the present paper.
Nevertheless it is very instructive, for a better understanding of these non-relativistic
results, to start from the original relativistic form of the RST dynamics and to glimpse
at the emergence of its non-relativistic approximation which will then afterwards be
inspected and treated in great detail.

\begin{center}
  \emph{\textbf{1.\ Hamilton-Lagrange Action Principle}}
\end{center}

Similarly to most of the modern particle theories, the RST dynamics can also be based
upon an action principle~\cite{SSMS}
\begin{equation}
  \label{eq:II.1}
  \delta\WRST=0\ ,
\end{equation}
where the action integral~$\WRST$ refers to some Lagrangean~$\LRST[\Psi,\A_\mu]$
\begin{equation}
  \label{eq:II.2}
  \WRST=\int\! d^4 x\;\LRST[\Psi,\A_\mu]\ .
\end{equation}
This Lagrangean appears as the sum of a matter part~$\LD$ and a gauge field part~$\LG$
\begin{equation}
  \label{eq:II.3}
  \LRST[\Psi,\A_\mu]=\LD[\Psi]+\LG[\A_\mu].
\end{equation}
The matter part~$\LD$ is given in terms of the wave function~$\Psi$ by~\cite{MaSo}
\begin{equation}
  \label{eq:II.4}
  \LD[\Psi] = \frac{i\hbar c}{2}\left[\bar{\Psi}\GG^\mu\left(\D_\mu\Psi\right) -
 \left(\D_\mu\bar{\Psi}\right)\GG^\mu\Psi \right] - \bar{\Psi}\M c^2\Psi \ ,
\end{equation}
and similarly the gauge field part~$\LG$ reads in terms of the bundle
curvature~$\F_{\mu\nu}$ (``\emph{field strength}'' )
\begin{gather}
  \label{eq:II.5}
  \LG[\A_\mu]=\frac{\hbar c}{16\pi\as}K_{\alpha\beta}{F^\alpha}_{\mu\nu} F^{\beta\mu\nu}\\*
\left( \as\doteqdot\frac{e^2}{\hbar c}\right)\ .\notag
\end{gather}
The curvature components~${F^\alpha}_{\mu\nu}$ do refer here to the decomposition of the
bundle curvature~$\F_{\mu\nu}$
\begin{equation}
  \label{eq:II.6}
  \F_{\mu\nu} = \nabla_\mu\A_\nu - \nabla_\nu\A_\mu + [\A_\mu,\A_\nu]
\end{equation}
with respect to a suitably chosen basis~$\{\tau_\alpha \}$ of the  structure algebra
\begin{equation}
  \label{eq:II.7}
  \F_{\mu\nu} = {F^\alpha}_{\mu\nu}\tau_\alpha\ .
\end{equation}
This basis~$\{\tau_\alpha\}$ builds up also the metric~$K_{\alpha\beta}$ for the Lie algebra bundle in the
following way:
\begin{equation}
  \label{eq:II.8}
  K_{\alpha\beta} =
  c_1\tr(\tau_\alpha)\cdot\tr(\tau_\beta)+c_2\tr(\tau_\alpha\cdot\tau_\beta)\ .
\end{equation}
Here the coefficients~$c_1$ and~$c_2$ are ordinary constants over space-time so
that~$K_{\alpha\beta}$ is actually covariantly constant
\begin{equation}
  \label{eq:II.9}
  D_\mu K_{\alpha\beta} \equiv 0\ .
\end{equation}

The gauge forces among the particles are incorporated into this formalism via the principle
of \emph{minimal coupling}, i.e.\ the covariant derivative of the wave function~$\Psi$ in
the matter Lagrangean~$\LD$ (\ref{eq:II.4}) is defined by
\begin{equation}
  \label{eq:II.10}
  \D_\mu\Psi = \partial_\mu\Psi + \A_\mu\Psi\ .
\end{equation}
Here the bundle connection~$\A_\mu$ (``\emph{gauge potential}'') can also be decomposed
with respect to the Lie algebra basis~$\{\tau_\alpha\}$, quite analogously to its
curvature~$\F_{\mu\nu}$ (\ref{eq:II.7})
\begin{equation}
  \label{eq:II.11}
  \A_\mu = {A^\alpha}_\mu\tau_\alpha\ ,
\end{equation}
i.e.\ for a two particle system whose gauge algebra turns out to be
four-dimensional~$(\alpha=1\ldots 4)$ with~$\tau_3\doteqdot\chi, \tau_4\doteqdot\bar{\chi}$:
\begin{subequations}
  \begin{align}
  \label{eq:II.12a}
  \A_\mu &= {A^1}_\mu\tau_1 + {A^2}_\mu\tau_2 + B_\mu\chi - \Bstar_\mu\bar{\chi}\\*
  \label{eq:II.12b}
  \F_{\mu\nu} &=
  {F^1}_{\mu\nu}\tau_1+{F^2}_{\mu\nu}\tau_2+G_{\mu\nu}\chi-\Gstar_{\mu\nu}\bar{\chi}\ .
  \end{align}
\end{subequations}

These decompositions of the gauge potential and field strength demonstrate that the
totality of gauge forces in RST subdivide into two classes: \\
\textbf{(i)} \emph{electromagnetic forces} being described by the real-valued
four-potentials ${A^a}_\mu$ and field strengths~${F^a}_{\mu\nu}\,(a=1,2)$ and \textbf{(ii)}
\emph{exchange forces} being described by the complex-valued four-potential~$B_\mu$.
However the point with the exchange type of interaction is that this kind of force can be
active exclusively among \emph{identical} particles~\cite{MaSo,BS} while in the present paper
the interest aims at positronium as a bound state of two \emph{different} particles, i.e.\
positron and electron. Therefore, for such a two-particle system, the exchange
potential~$B_\mu$ and its field strength~$G_{\mu\nu}$ must be put to zero~$(B_\mu\equiv 0,
G_{\mu\nu} \equiv 0)$ so that the gauge fields (\ref{eq:II.12a})-(\ref{eq:II.12b}) are
reduced to their Abelian projections~$(a=1,2)$
\begin{subequations}
  \begin{align}
    \label{eq:II.13a}
    \A_\mu&\Rightarrow {A^a}_\mu\tau_a\\*
    \label{eq:II.13b}
    \F_{\mu\nu}&\Rightarrow{F^a}_{\mu\nu}\tau_a\ ,
  \end{align}
\end{subequations}
with the usual link of potentials and field strengths
\begin{equation}
  \label{eq:II.14}
  {F^a}_{\mu\nu} = \nabla_\mu{A^a}_\nu -\nabla_\nu{A^a}_\mu\ .
\end{equation}

A similar simplification must occur also for the matter fields because the material
particles (i.e.\ electron and positron) cannot feel the exchange forces. More precisely,
since the two-particle wave function~$\Psi$ is the Whitney sum of the two single-particle
wave functions~$\psi_a\,(a=1,2)$, i.e.
\begin{equation}
  \label{eq:II.15}
  \Psi(x) = \psi_1(x)\oplus\psi_2(x)\ ,
\end{equation}
the covariant derivative of $\Psi$ appears as
\begin{equation}
  \label{eq:II.16}
  \D_\mu\Psi = D_\mu\psi_1 \oplus D_\mu\psi_2
\end{equation}
with
\begin{subequations}
  \begin{align}
    \label{eq:II.17a}
    D_\mu\psi_1 &= \partial_\mu\psi_1 - i{A^2}_\mu\psi_1\\*
    \label{eq:II.17b}
    D_\mu\psi_2 &= \partial_\mu\psi_2 - i{A^1}_\mu\psi_2\ .
  \end{align}
\end{subequations}

Finally, it should be self-evident that the missing of the exchange forces does also
simplify the fibre metric~$K_{\alpha\beta}$ (\ref{eq:II.8}) since this missing 
reduces the~$(4\times 4)$-matrix~$K_{\alpha\beta}$ to the
following~$(2\times 2)$-matrix~$K_{ab}$
\begin{equation}
  \label{eq:II.18}
  \left\{K_{\alpha\beta}\right\} \Rightarrow \left\{K_{ab}\right\} =
  \begin{pmatrix}
    0 & -1 \\ -1 & 0 \\
  \end{pmatrix}\ ,
\end{equation}
which obviously then simplifies the gauge field Lagrangean~$\LG$ (\ref{eq:II.5}) to
\begin{equation}
  \label{eq:II.19}
  \LG[\A_\mu] \Rightarrow - \frac{\hbar c}{8\pi\as}{F^1}_{\mu\nu} F^{2\mu\nu}\ .
\end{equation}
\pagebreak
\begin{center}
  \emph{\textbf{2.\ Field Equations and Conservation Laws}}
\end{center}

Once the kinematical setting for the action integral~$\WRST$ (\ref{eq:II.2}) is
established, the dynamical equations of the theory do appear as the corresponding
Euler-Lagrange equations; namely, the variation (\ref{eq:II.1}) with respect to the matter
field~$\Psi$ yields the Dirac equation (DE)
\begin{equation}
  \label{eq:II.20}
  i\hbar\GG^\mu\D_\mu\Psi=\M c\Psi\ ,
\end{equation}
or, resp., in component form for the presently considered two-particle systems~\cite{BMS}
\begin{subequations}
  \begin{align}
    \label{eq:II.21a}
    i\hbar\gamma^\mu D_\mu\psi_1 &= -\Mp c \psi_1\\*
    \label{eq:II.21b}
    i\hbar\gamma^\mu D_\mu\psi_2 &= \Me c \psi_2
  \end{align}
\end{subequations}
where~$\Mp (\Me)$ is the rest mass of the positively (negatively) charged particle. (For
the positronium system, both masses are identical:~$\Mp=\Me\doteqdot M$). Quite
analogously, the variation (\ref{eq:II.1}) with respect to the bundle connection~$\A_\mu$
yields the (non-Abelian) Maxwell equations
\begin{equation}
  \label{eq:II.22}
  \D^\mu\F_{\mu\nu} = -4\pi i \as\J_\nu\ ,
\end{equation}
or, resp., in component form
\begin{gather}
  \label{eq:II.23}
  D^\mu {F^\alpha}_{\mu\nu} = 4\pi\as{j^\alpha}_\nu \\*
  \left(\alpha=1\ldots 4\right)\ .\notag
\end{gather}
But the point with these general gauge field equations is now that they simplify to the
Abelian form
\begin{gather}
  \label{eq:II.24}
  \nabla^\mu {F^a}_{\mu\nu} = 4\pi\as{j^a}_\nu\\*
  (a=1,2)\ ,\notag
\end{gather}
because the absence of the exchange interactions for non-identical particles~$(\leadsto
B_\mu \equiv 0)$ eliminates the non-Abelian part of the gauge field equations, see
ref.s~\cite{MaSo,BMS}. 

The coupled set of matter and gauge field equations (\ref{eq:II.20}) and (\ref{eq:II.22})
is not yet a closed system because it is necessary to specify the \emph{Maxwell
  currents}~${j^\alpha}_\mu$ as the components of the current operator~$\J_\mu$
(\ref{eq:II.22})
\begin{equation}
  \label{eq:II.25}
  \J_\mu\doteqdot i{j^\alpha}_\mu\tau_\alpha\equiv i\left({j^1}_\mu\tau_1 +
    {j^2}_\mu\tau_2 + g_\mu\chi - \gstar_\mu\bar{\chi} \right)\ ,
\end{equation}
namely just in terms of the wave function~$\Psi$. This requirement, however, can be
satisfied by first linking the Maxwell currents~${j^\alpha}_\mu$ to the \emph{RST
  currents}~$j_{\alpha\mu}$ by means of the fibre metric~$K_{\alpha\beta}$ (\ref{eq:II.8})
\begin{subequations}
  \begin{align}
    \label{eq:II.26a}
    {j^\alpha}_\mu &= K^{\alpha\beta} j_{\beta\mu}\\*
    \label{eq:II.26b}
    j_{\alpha_\mu} &= K_{\alpha\beta} {j^\beta}_\mu\\*
    \left( K_{\alpha\beta}\right.  &K^{\beta\gamma} = \left.{\delta^\gamma}_\alpha\right)\ .\notag
  \end{align}
\end{subequations}
The point with this construction is namely that the RST currents~$j_{\alpha\mu}$ are
defined in terms of the wave function~$\Psi$ through
\begin{equation}
  \label{eq:II.27}
  j_{\alpha\mu} = \bar{\Psi}\GG_\mu\tau_\alpha\Psi\ ,
\end{equation}
and therefore these currents~$j_{\alpha\mu}$ do automatically obey the source equations
\begin{equation}
  \label{eq:II.28}
  D^\mu j_{\alpha\mu}\equiv 0
\end{equation}
which therefore appear as an immediate implication of the matter dynamics. On the other
hand, the covariant constancy of the fibre metric~$K_{\alpha\beta}$ (\ref{eq:II.9})
guarantees that also the Maxwell equations~${j^\alpha}_\mu$ do obey a source equation,
i.e.
\begin{equation}
  \label{eq:II.29}
  D^\mu {j^\alpha}_\mu\equiv 0\ ,
\end{equation}
which is nothing else than the component form of the abstract version
\begin{equation}
  \label{eq:II.30}
  \D^\mu\J_\mu\equiv 0\ .
\end{equation}
This identity, however, is itself an immediate implication of the gauge field equations
(\ref{eq:II.22}) so that the matter and gauge field dynamics are perfectly compatible with
each other.

Moreover, the absence of the exchange interactions for the positronium system reduces also
the source equation (\ref{eq:II.29}) to its Abelian truncation $(a=1,2)$
\begin{equation}
  \label{eq:II.31}
  \nabla^\mu {j^a}_\mu \equiv 0
\end{equation}
which itself is again an indispensable consistency condition for the Abelian
simplification (\ref{eq:II.24}) of the general Maxwell equation (\ref{eq:II.22}). Resorting
here to the usual Lorentz gauge condition for the electromagnetic potentials~${A^a}_\mu$
(\ref{eq:II.13a}) 
\begin{equation}
  \label{eq:II.32}
  \nabla^\mu {A^a}_\mu\equiv 0\ ,
\end{equation}
with observation of the curl relation (\ref{eq:II.14}), converts those Abelian Maxwell
equations (\ref{eq:II.24}) to the \emph{d'Alembert equations} for the
potentials~${A^a}_\mu$
\begin{equation}
  \label{eq:II.33}
  \partial^\mu\partial_\mu {A^a}_\nu = 4\pi\as{j^a}_\nu\ .
\end{equation}
Thus the RST dynamics is revealed as a consistently coupled set of matter and gauge field
equations, where the coupling of the gauge fields \mbox{${A^a}_\mu\, (a=1,2)$} to the
single-particle wave functions~$\psi_a$ can be made somewhat more obvious by introducing
the \emph{Dirac currents}~$k_{a\mu}$ in the usual way
\begin{equation}
  \label{eq:II.34}
  k_{a\mu}\doteqdot \bar{\psi}_a\gamma_\mu\psi_a\ .
\end{equation}
All three types of currents~$\{{j^a}_\mu,j_{a\mu},k_{a\mu}\}$ are then related to each
other by
\begin{subequations}
  \begin{align}
    \label{eq:II.35a}
    {j^1}_\mu &= -j_{2\mu} = k_{1\mu}\\*
    \label{eq:II.35b}
    {j^2}_\mu &= -j_{1\mu} = -k_{2\mu}
  \end{align}
\end{subequations}
and appear as being correctly matched to the positive charge of the first particle~($a=1$,
positron) and to the negative charge carried by the second particle~($a=2$, electron).

The conservation laws, such as those for the electric charges (\ref{eq:II.31}), are
relevant also in another respect: For a closed system with total energy-momentum
tensor~$\TT_{\mu\nu}$ one expects the validity of the conservation law
\begin{equation}
  \label{eq:II.36}
  \nabla^\mu\,\TT_{\mu\nu}\equiv 0\ .
\end{equation}
For a field theory which is based upon an action principle (\ref{eq:II.1}), such an
energy-momentum conservation law like (\ref{eq:II.36}) must emerge via the well-known
Noether theorems~\cite{SSMS}; or alternatively it must be possible to deduce this also
directly from the underlying dynamics (\ref{eq:II.20}) and
(\ref{eq:II.22})~\cite{GSMS}. Since the RST field system is built up by a matter part and a
gauge field part, it should appear natural that the total tensor~$\TT_{\mu\nu}$ turns out
as the sum of the matter density~$(\DT_{\mu\nu})$ and the gauge field
density~$(\GT_{\mu\nu})$, i.e.
\begin{equation}
  \label{eq:II.37}
  \TT_{\mu\nu} =   \DT_{\mu\nu} +   \GT_{\mu\nu}\ ,
\end{equation}
with the matter part being given by 
\begin{equation}
  \label{eq:II.38}
  \DT_{\mu\nu} = \frac{i\hbar c}{4}\left[ \bar{\Psi}\GG_\mu \left(\D_\nu\Psi\right)-
\left(\D_\nu\bar{\Psi}\right)\GG_\mu\Psi+\bar{\Psi}\GG_\nu \left(\D_\mu\Psi\right) -
\left(\D_\mu\bar{\Psi}\right)\GG_\nu\Psi\right]
\end{equation}
and the gauge field part by
\begin{equation}
  \label{eq:II.39}
  \GT_{\mu\nu} = \frac{\hbar c}{4\pi\as}\,K_{\alpha\beta} \left(
    {F^\alpha}_{\mu\lambda}{F^\beta}_\nu{}^\lambda - \frac{1}{4} g_{\mu\nu}
    {F^\alpha}_{\lambda\sigma}F^{\beta\lambda\sigma}\right)\ .
\end{equation}
By means of the dynamical equations for the matter fields (\ref{eq:II.20}) and gauge
fields (\ref{eq:II.22}) one can now show that the conservation law (\ref{eq:II.36}) is
actually obeyed because the sources of the partial densities~$\DT_{\mu\nu}$
and~$\GT_{\mu\nu}$ do just cancel
\begin{equation}
  \label{eq:II.40}
   \nabla^\mu\,\DT_{\mu\nu} = - \nabla^\mu\,\GT_{\mu\nu} = \hbar
   c{F^\alpha}_{\mu\nu}{j_\alpha}^\mu\ .
\end{equation}
As we shall readily see, the existence of such a total density~$\TT_{\mu\nu}$ is important
for the subsequent construction of the \emph{principle of minimal energy} which will be
based upon the total energy density~$\TT_{00}(\vec{r})$ as the time-component
of~$\TT_{\mu\nu}$.

\begin{center}
  \emph{\textbf{3.\ Stationary Bound States}}
\end{center}

When the positronium system has settled down to its groundstate, the corresponding field
configurations must be expected to be stationary or even time-independent. For the wave
functions~$\psi_a(\vec{r},t)$ this implies the following product form~$(a=1,2)$
\begin{equation}
  \label{eq:II.41}
  \psi_a(\vec{r},t) = \exp\left(-i\,\frac{M_a c^2}{\hbar}\,t \right)\cdot\psi_a(\vec{r})\ ,
\end{equation}
while the currents and gauge potentials will be found to be time-independent:
\begin{subequations}
  \begin{align}
    \label{eq:II.42a}
    k_{a\mu}(\vec{r},t) &\Rightarrow \{\akn,-\vec{k}_a(\vec{r}) \}\\*
    \label{eq:II.42b}
    {A^a}_\mu(\vec{r},t) &\Rightarrow \{\aAn; -\vec{A}_a(\vec{r}) \}\ .
  \end{align}
\end{subequations}
Accordingly, the d'Alembert equations (\ref{eq:II.33}) are reduced to the
ordinary \emph{Poisson equations}
\begin{subequations}
  \begin{align}
    \label{eq:II.43a}
    \Delta\iAn &= -4\pi\as\ikn\\*
    \label{eq:II.43b}
    \Delta\iiAn &= 4\pi\as\iikn\\*
    \label{eq:II.43c}
    \Delta\vec{A}_1(\vec{r}) &= -4\pi\as\vec{k}_1(\vec{r})\\*
    \label{eq:II.43d}
    \Delta\vec{A}_2(\vec{r}) &= 4\pi\as\vec{k}_2(\vec{r})
  \end{align}
\end{subequations}
whose formal solutions are given by
\begin{subequations}
  \begin{align}
    \label{eq:II.44a}
    \fAn &= \pm\as\int d^3\vec{r}\,'\frac{\sakn}{||\vec{r}-\vec{r}\,'||}\\*
    \label{eq:II.44b}
    \vec{A}_{1/2}(\vec{r}) &= \pm\as\int d^3\vec{r}\,'
    \frac{\vec{k}_a(\vec{r}\,')}{||\vec{r}-\vec{r}\,'||}\ .
  \end{align}
\end{subequations}

Similarly, the time-independent \emph{mass eigenvalue equations} for the spatial parts of
the wave functions~$\psi_a(\vec{r})$ (\ref{eq:II.41}) may be deduced directly from the
matter equations (\ref{eq:II.21a})-(\ref{eq:II.21b}), or they can be deduced alternatively
by substituting the stationary ansatz (\ref{eq:II.41}) into the action principle
(\ref{eq:II.1})-(\ref{eq:II.2}) and carrying through the variational procedure. The result
is then found to be of the following form
\begin{subequations}
  \begin{align}
    \label{eq:II.45a}
    i\vec{\gamma}\sdot\vec{\nabla}\psi_1(\vec{r}) + \iiAn\gamma^0\psi_1(\vec{r}) -
      \vec{A}_2(\vec{r})\sdot\vec{\gamma}\psi_1(\vec{r}) &=
      -\left(\frac{M_1 c}{\hbar}\gamma_0+\frac{\Mp c}{\hbar} \right)\psi_1(\vec{r})\\*
    \label{eq:II.45b}
    i\vec{\gamma}\sdot\vec{\nabla}\psi_2(\vec{r}) + \iAn\gamma^0\psi_2(\vec{r}) -
      \vec{A}_1(\vec{r})\sdot\vec{\gamma}\psi_2(\vec{r}) &=
      -\left(\frac{M_2 c}{\hbar}\gamma_0-\frac{\Me c}{\hbar} \right)\psi_2(\vec{r})\ .
  \end{align}
\end{subequations}
For obtaining here the solutions~$\psi_a(\vec{r})$, it is very convenient to conceive the
Dirac four-spinors~$\psi_a(\vec{r})$ as the (local) direct sum of the Pauli
two-spinors $\ap$:
\begin{equation}
  \label{eq:II.46}
  \psi_a(\vec{r}) = \app\oplus\apm\ ,
\end{equation}
and for the subsequent treatment of the positronium groundstate one decomposes these Pauli
spinors with respect to a certain basis system~$\{\onpm,\oipm\}$ as
\begin{subequations}
  \begin{align}
    \label{eq:II.47a}
    \app &= (r\sin\vartheta)^{-\frac{1}{2}}\left(\aRp\rt\cdot\onp +
      \aSp\rt \cdot \onm \right)\\*
    \label{eq:II.47b}
    \apm &= -i (r\sin\vartheta)^{-\frac{1}{2}}\left(\aRm\rt\cdot\oip +
      \aSm\rt \cdot \oim \right)    \ .
  \end{align}
\end{subequations}
(For the details see ref.~\cite{BMS}). Clearly by such an arrangement, the original mass
eigenvalue equations (\ref{eq:II.45a})-(\ref{eq:II.45b}) transcribe to the corresponding
equations for the (real-valued) \emph{wave amplitudes}~$\aRpm,\aSpm$; e.g.\ for the first
particle~($a=1$, positron):
\begin{subequations}
  \begin{align}
    \label{eq:II.48a}
    \begin{split}
    \frac{\partial\iRp}{\partial r}+\frac{1}{r}\frac{\partial\iSp}{\partial\vartheta}
    +{}^{(2)}\!A_0\cdot\iRm-{}^{(2)}\!A_\phi\left(\sin\vartheta\cdot\iRp-\cos\vartheta\cdot\iSp\right)\\
    = \frac{\Mp-M_1}{\hbar}\,c\cdot\iRm  
    \end{split}\\
    \label{eq:II.48b}
    \begin{split}
    \frac{\partial\iSp}{\partial r}-\frac{1}{r}\frac{\partial\iRp}{\partial\vartheta}
    +{}^{(2)}\!A_0\cdot\iSm+{}^{(2)}\!A_\phi\left(\sin\vartheta\cdot\iSp+\cos\vartheta\cdot\iRp\right)\\
    = \frac{\Mp-M_1}{\hbar}\,c\cdot\iSm
    \end{split}\\
    \label{eq:II.48c}
    \begin{split}
    \frac{1}{r}\frac{\partial(r\iRm)}{\partial r}-\frac{1}{r}\frac{\partial\iSm}{\partial\vartheta}
    -{}^{(2)}\!A_0\cdot\iRp+{}^{(2)}\!A_\phi\left(\sin\vartheta\cdot\iRm-\cos\vartheta\cdot\iSm\right)\\
    = \frac{\Mp+M_1}{\hbar}\,c\cdot\iRp
    \end{split}\\
    \label{eq:II.48d}
    \begin{split}
    \frac{1}{r}\frac{\partial(r\iSm)}{\partial r}+\frac{1}{r}\frac{\partial\iRm}{\partial\vartheta}
    -{}^{(2)}\!A_0\cdot\iSp-{}^{(2)}\!A_\phi\left(\sin\vartheta\cdot\iSm+\cos\vartheta\cdot\iRm\right)\\
    = \frac{\Mp+M_1}{\hbar}\,c\cdot\iSp\ .      
    \end{split}
  \end{align}
\end{subequations}

Here it can be shown that for real-valued wave amplitudes~$\aRpm,\aSpm$ the Dirac
three-currents~$\vec{k}_a(\vec{r})$ (\ref{eq:II.42a}) encircle the
z-axis~$(\vartheta=0,\pi)$ of the spherical polar coordinates~$(r,\vartheta,\phi)$, i.e.
\begin{equation}
  \label{eq:II.49}
  \vec{k}_a(\vec{r}) = {}^{(a)}k_\phi(r,\vartheta)\cdot\vec{e}_\phi\ ,
\end{equation}
and therefore it is self-suggesting to adopt a similar symmetry for the magnetic
potentials~$\vec{A}_a(\vec{r})$ (\ref{eq:II.42b}):
\begin{equation}
  \label{eq:II.50}
  \vec{A}_a(\vec{r}) = {}^{(a)}\!A_\phi(r,\vartheta)\cdot\vec{e}_\phi\ .
\end{equation}
The magnetic Poisson equations  (\ref{eq:II.43c})-(\ref{eq:II.43d}) are then transcribed
to the azimuthal components~${}^{(a)}\!A_\phi$ of these vector
potentials~$\vec{A}_a(\vec{r})$ as
\begin{equation}
  \label{eq:II.51}
  \Delta {}^{(1/2)}\!A_\phi - \frac{{}^{(1/2)}\!A_\phi}{r^2\sin^2\vartheta} =
  \mp 4\pi\as {}^{(1/2)} k_\phi\ .
\end{equation}

\begin{center}
  \emph{\textbf{4.\ Non-Relativistic Approximation}}
\end{center}

For the subsequent numerical demonstrations, we will restrict ourselves to the
non-relativistic approximation; and for this purpose one therefore has to elaborate now
the non-relativistic limit form of both the electrostatic Poisson equations
(\ref{eq:II.43a})-(\ref{eq:II.43b}) and the mass eigenvalue equations
(\ref{eq:II.48a})-(\ref{eq:II.48d}). For the case of the Poisson equations, this problem
is much simpler than for the eigenvalue equations, because the general form of the Poisson
equations survives the non-relativistic limit so that it becomes merely necessary to
specify the non-relativistic approximation of the Dirac densities~$\akn$
(\ref{eq:II.42a}). But here one can easily show, see ref.~\cite{BeSo}, that the desired
non-relativistic form of these charge densities is obtained by simply neglecting the
``negative'' Pauli components~$\apm$ (\ref{eq:II.47b}) and by additionally omitting
alternatively one of the two spin directions. This means that the spin of any particle
points to a definite z-direction, either the positive one~$(\aSp\equiv 0)$ or the negative
one~$(\aRp\equiv 0)$; and thus both spin orientations become decoupled in the sense that
the wave amplitudes do obey separate non-relativistic eigenvalue equations, e.g.\ for the
first particle (with positive z-component of spin, say):
\begin{gather}
  \label{eq:II.52}
  \begin{split}    
  -\frac{\hbar^2}{2\Mp}\left[\frac{1}{r}\frac{\partial}{\partial r} \left( r\cdot
      \frac{\partial \iRp}{\partial r}\right) + \frac{1}{r^2}
      \frac{\partial^2}{\partial \vartheta^2}\, \iRp \right] +
    \hbar c\, {}^{(2)}\!A_0 \rt\cdot\iRp\\ = \ES_{(1)}\cdot\iRp
  \end{split}\\
  \left( \ES_{(1)}\doteqdot-\left(\Mp+M_1\right)c^2 \right)\notag
\end{gather}
or analogously for the second particle (with negative z-component of spin, say)
\begin{gather}
  \label{eq:II.53}
  \begin{split}   
  -\frac{\hbar^2}{2\Me}\left[\frac{1}{r}\frac{\partial}{\partial r} \left( r\cdot
      \frac{\partial \iiSp}{\partial r}  \right) + \frac{1}{r^2} \frac{\partial^2}{\partial \vartheta^2}\, \iiSp \right) -
    \hbar c\,{}^{(1)}\!A_0 \rt\cdot\iiSp\\ = \ES_{(2)}\cdot\iiSp
  \end{split}\\
  \left( \ES_{(2)}\doteqdot \left(M_2-\Me\right)c^2\right)\ .  \notag
\end{gather}
However, observe here that these non-relativsitic energy eigenvalue equations are not of
the usual Schr\"odinger form! This is due to the fact that the eigenvalue of the total
angular momentum~$(\hat{J}_z)$ is zero (see the discussion of this effect in
ref.~\cite{BMS}). But on the other hand the non-relativistic energy
eigenvalues~$\ES_{(a)}$ are linked here as usual to the difference of rest
masses~$M_{\mathrm{p/e}}$ and mass eigenvalues~$M_a$ as shown above.

Concerning now the non-relativistic approximation of the electrostatic Poisson equations
(\ref{eq:II.43a})-(\ref{eq:II.43b}) one finds for the Dirac densities~$\akn$~\cite{BeSo}
\begin{subequations}
  \begin{align}
    \label{eq:II.54a}
    {{}^{(1)}\!k_0(r,\vartheta)} &\simeq \frac{\iRp^2(r,\vartheta)}{4\pi r\sin\vartheta}\\*[5mm]
    \label{eq:II.54b}
    {{}^{(2)}\!k_0(r,\vartheta)} &\simeq \frac{\iiSp^2(r,\vartheta)}{4\pi
      r\sin\vartheta}\ ,
  \end{align}
\end{subequations}
provided one adopts again the first (second) spin pointing to the positive (negative)
z-direction. Thus the electric Poisson equations (\ref{eq:II.43a})-(\ref{eq:II.43b}) appear
as
\begin{subequations}
  \begin{align}
    \label{eq:II.55a}
    \Delta\,{}^{(1)}\!\!A_0(r,\vartheta) &= -\as\cdot\frac{\iRp^2(r,\vartheta)}{r\sin\vartheta}\\*[5mm]
    \label{eq:II.55b}
    \Delta\,{}^{(2)}\!\!A_0(r,\vartheta) &= \as\cdot\frac{\iiSp^2(r,\vartheta)}{r\sin\vartheta}\ .
  \end{align}
\end{subequations}
In contrast to their \emph{electrostatic} counterparts, the \emph{magnetostatic} Poisson
equations (\ref{eq:II.43c})-(\ref{eq:II.43d}) are disregarded because the magnetic effects
appear to be of the same order of magnitude as the relativistic effects, which of course
are to be neglected for the present non-relativistic approximation. This is also the
reason why the magnetic potentials~${}^{(a)}\!A_\phi(r,\vartheta)$ have been omitted for
the non-relativistic approximation of the mass eigenvalue equations, whereas the original
relativistic form (\ref{eq:II.48a})-(\ref{eq:II.48d}) must of course contain the magnetic
potentials~${}^{(a)}\!A_\phi(r,\vartheta)$ which, e.g., describe the spin-spin
interactions for the groundstate.

However, the crucial point with the residual electric Poisson equations
(\ref{eq:II.55a})-(\ref{eq:II.55b}) is now that they obviously will yield \emph{angular
  dependent} potentials~${}^{(a)}\!A_0(r,\vartheta)$, even if one adopts the wave
amplitudes~$\iRp,\iiSp$ to be spherically symmetric! Clearly, these angular dependent
potentials~${}^{(a)}\!A_0(r,\vartheta)$ as solutions of the Poisson equations
(\ref{eq:II.55a})-(\ref{eq:II.55b}) must then entail their angular dependence also on the
wave amplitudes~$\iRp,\iiSp$ as solutions of the (non-relativistic) energy eigenvalue
equations (\ref{eq:II.52})-(\ref{eq:II.53}). But this angular dependence of the RST
eigenvalue equations will be neglected subsequently by resorting to the
``\emph{spherically symmetric approximation}''~\cite{BMS}. Namely, by this approximation do
our \emph{non-relativistic} RST predictions stand on the same footing as the conventional
treatment of positronium which relies upon the spherical symmetry of the Coulomb potential
for the internal two-body problem (see the corresponding remarks in the \emph{Introduction}).
Here, the neglection of the anisotropic effect means for the (non-relativistic) energy
eigenvalue equations (\ref{eq:II.52})-(\ref{eq:II.53}) that one simply puts to zero the
angular derivatives of the wave amplitudes~$\iRp$ and~$\iiSp$; and furthermore one adopts
also the spherical symmetry for the electric potentials which then recasts those
eigenvalue equations to the following truncated form:
\begin{subequations}
  \begin{align}
\label{eq:II.56a}
-\frac{\hbar^2}{2\Mp}\cdot \frac{1}{r}\frac{d}{dr}\left(r\cdot \frac{d\,\iRp(r)}{dr}
\right)+\hbar c\, {}^{[2]}\!A_0(r)\cdot\iRp(r) &= \ES_{(1)}\cdot\iRp(r)\\*
\label{eq:II.56b}
-\frac{\hbar^2}{2\Me}\cdot \frac{1}{r}\frac{d}{dr}\left(r\cdot \frac{d\,\iiSp(r)}{dr}
\right)-\hbar c\, {}^{[1]}\!A_0(r)\cdot\iiSp(r) &= \ES_{(2)}\cdot\iiSp(r)\ .
  \end{align}
\end{subequations}
Observe here that, in contrast to the situation with these spherically symmetric
eigenvalue equations, the corresponding symmetric approximations~${}^{[a]}\!A_0(r)$ of the
electric potentials~${}^{(a)}\!A_0(r,\vartheta)$ cannot be obtained by simply omitting the
angular derivatives in their Poisson equations (\ref{eq:II.55a})-(\ref{eq:II.55b}), but
the desired spherically symmetric Poisson equations must be deduced \emph{directly} from
the principle of minimal energy which is now to be considered in greater detail.


\section{Principle of Minimal Energy}
\indent

As pleasant as the existence of the action principle (\ref{eq:II.1}) for RST may appear,
the action integral (\ref{eq:II.2}) itself is not equipped with an immediate physical
meaning. In this context, it seems highly desirable to have some variational principle
which does refer to the \emph{total field energy}~$\ET$ being stored in the RST field
configurations. If this goal could be attained, both the mass eigenvalue equations and the
Poisson equations would \emph{not only} appear as the (stationary) Euler-Lagrange
equations due to the Hamilton-Lagrange action principle (\ref{eq:II.1})-(\ref{eq:II.2})
but this eigenvalue problem would then appear also in form of the variational equations
due to that total energy functional~$\ET$. Possibly, the variations of the
functional~$\ET$ must be restricted by some constraints. Fortunately, it has already been
demonstrated that the construction of such an energy functional~$\ET$ is possible, for
both the relativistic and the non-relativistic situation~\cite{MaSo,BMS}.

\begin{center}
  \emph{\textbf{1.\ Relativistic Construction}}
\end{center}

Naturally, one will assume that the wanted energy functional~$(\tET)$ will be related in one
way or the other to the total energy density~$\ET$ as the spatial integral over the
total energy density~$\TT_{00}(\vec{r})$:
\begin{subequations}
  \begin{align}
    \label{eq:III.1a}
    \ET &= \int d^3\vec{r}\;\TT_{00}(\vec{r}) = \ED + \EG\\*
    \label{eq:III.1b}
    \ED &= \int d^3\vec{r}\;\DT_{00}(\vec{r})\\*
    \label{eq:III.1c}
    \EG &= \int d^3\vec{r}\;\GT_{00}(\vec{r})\ .
  \end{align}
\end{subequations}
Clearly, since the total energy-momentum density~$\TT_{\mu\nu}$ (\ref{eq:II.37}) appears
as the sum of the matter part~$\DT_{\mu\nu}$ (\ref{eq:II.38}) and of the gauge
part~$\GT_{\mu\nu}$ (\ref{eq:II.39}) the total energy~$\ET$ is the sum of the matter
energy~$\ED$ and the gauge field energy~$\EG$.

Furthermore, since the two-particle wave-function~$\Psi(\vec{r})$ in the matter
density~$\DT_{\mu\nu}$ (\ref{eq:II.38}) is the Whitney sum (\ref{eq:II.15}) of the two
one-particle functions~$\psi_a(\vec{r})$, the matter energy~$\ED$ (\ref{eq:III.1b}) is
revealed as the sum of the individual energies~$\ED_{(a)}$ of both particles:
\begin{equation}
  \label{eq:III.2}
  \ED = \ED_{(1)} + \ED_{(2)}\ ,
\end{equation}
with the one-particle contributions~$\ED_{(a)}$ being given by~\cite{MaSo}
\begin{subequations}
  \begin{align}
    \label{eq:III.3a}
    \ED_{(1)} &= \Z_{(1)}^2\cdot\Mp c^2 + 2\Tkin_{(1)} + \Mmi c^2\\*
    \label{eq:III.3b}
    \ED_{(2)} &= \Z_{(2)}^2\cdot\Me c^2 + 2\Tkin_{(2)} + \Mmii c^2\ .
  \end{align}
\end{subequations}
Surely, this is a very plausible result because it says that the material
energies~$\ED_{(a)}$ of both particles consist of: 
\begin{itemize}
\item[\textbf{(i)}] the rest mass energies~$\Mpe c^2$, to be corrected by the \emph{mass
    renormalization factors}~$\Z_{(a)}$
  \begin{equation}
    \label{eq:III.4}
    \Z_{(a)}^2 = \int d^3\vec{r}\;\bar{\psi}_a(\vec{r})\psi_a(\vec{r})
  \end{equation}
\item[\textbf{(ii)}] the proper kinetic energies~$\Tkin_{(a)}$
  \begin{subequations}
    \begin{align}
      \label{eq:III.5a}
      \Tkin_{(1)} &= \frac{i}{2}\hbar c\int d^3\vec{r}\;\bar{\psi}_1(\vec{r})\vec{\gamma}\sdot
      \vec{\nabla}\psi_1(\vec{r})\\*
      \label{eq:III.5b}
      \Tkin_{(2)} &= -\frac{i}{2}\hbar c\int d^3\vec{r}\;\bar{\psi}_2(\vec{r})\vec{\gamma}\sdot
      \vec{\nabla}\psi_2(\vec{r})
    \end{align}
  \end{subequations}
    (for the factor of two in front of the kinetic energies~$\Tkin_{(a)}$
    (\ref{eq:III.3a})-(\ref{eq:III.3b}) see ref.~\cite{PS})
  \item[\textbf{(iii)}] the mass equivalents~$\Mmiii c^2$ of the magnetic interaction
    energy
    \begin{subequations}
      \begin{align}
        \label{eq:III.6a}
        \Mmi c^2 &= -\hbar c\int d^3\vec{r}\;\vec{A}_2(\vec{r})\sdot\vec{k}_1(\vec{r})\\*
        \label{eq:III.6b}
        \Mmii c^2 &= \hbar c\int d^3\vec{r}\;\vec{A}_1(\vec{r})\sdot\vec{k}_2(\vec{r})\ .
      \end{align}
    \end{subequations}
\end{itemize}

Observe here that, in contrast to the magnetic case, the mass equivalents~$\Meiii c^2$ of
the \emph{electric} interaction energy
\begin{subequations}
  \begin{align}
    \label{eq:III.7a}
    \Mei c^2 &\doteqdot \hbar c\int d^3\vec{r}\; \iiAn\cdot\ikn\\*
    \label{eq:III.7b}
    \Meii c^2 &\doteqdot -\hbar c\int d^3\vec{r}\; \iAn\cdot\iikn
  \end{align}
\end{subequations}
are not part of the matter energies~$\ED_{(a)}$ and thus there is here no perfect symmetry
between the electric and magnetic interactions. This symmetry, however, is manifestly
present in the gauge field~$\EG$ (\ref{eq:III.1c}) which does appear here as the
interaction energy~$\hER$ of the real gauge field modes, i.e.
\begin{equation}
  \label{eq:III.8}
  \EG \Rightarrow \hER = \heER + \hmER\ ,
\end{equation}
with the electric and magnetic parts being given by
\begin{subequations}
  \begin{align}
    \label{eq:III.9a}
    \heER &= \frac{\hbar c}{4\pi\as}\int d^3\vec{r}\;\vec{\nabla}\iAn\sdot\vec{\nabla}\iiAn\\*
    \label{eq:III.9b}
    \hmER &= \frac{\hbar c}{4\pi\as}\int d^3\vec{r}\;\vec{H}_1\sdot\vec{H}_2\\*
    \Big( &\vec{H}_a(\vec{r}) \doteqdot \vec{\nabla}\times\vec{A}_a(\vec{r}) \Big)\ . \notag
  \end{align}
\end{subequations}

But though, with these arrangements, the total field energy~$\ET$ (\ref{eq:III.1a}) is
well in agreement with the conventional field-theoretic conceptions, this functional~$\ET$
cannot serve directly for the wanted variational principle. The reason is that the
corresponding variational equations (due to~$\delta\ET=0$) do \emph{not} coincide with the
coupled set of Poisson and mass eigenvalue equations (\ref{eq:II.43a})-(\ref{eq:II.43d})
and (\ref{eq:II.45a})-(\ref{eq:II.45b})! The reason for this is mainly that the total
energy functional~$\ET$ (\ref{eq:III.1a})-(\ref{eq:III.1c}) contains no coupling of the
matter fields~$\psi_a(\vec{r})$ to the electric gauge potentials~$\aAn$, while such a
coupling is present in both the mass eigenvalue equations
(\ref{eq:II.45a})-(\ref{eq:II.45b}) and the Poisson equations
(\ref{eq:II.43a})-(\ref{eq:II.43d}). Consequently, it becomes now necessary to introduce
the missing couplings in such a way that, on the one hand, the numerical value of the
original functional~$\ET$ upon the solutions of the RST eigenvalue problem is left
unchanged but, on the other hand, the coupled eigenvalue and Poisson equations do actually
emerge as the variational equations due to the modified functional~$\tET$. This problem of
modifying the original~$\ET$ to~$\tET$ under preservation of its numerical value can
obviously be solved by imposing upon the variational process certain constraints which are
automatically obeyed by all the solutions of the RST eigenvalue problem. Here, the first
constraint refers to the fact that the mass eigenvalue equations
(\ref{eq:II.45a})-(\ref{eq:II.45b}) for fixed potentials~$\aAn,\vec{A}_a(\vec{r})$ are
formally linear which admits to normalize the solutions~$\psi_a(\vec{r})$ to unity~$(a=1,2)$:
\begin{equation}
  \label{eq:III.10}
  \int d^3\vec{r}\; \akn = \int d^3\vec{r}\;\psi^\dagger_a(\vec{r})\psi_a(\vec{r})=1\ .
\end{equation}
Thus the first constraint is that of wave function normalization and reads as
follows:
\begin{equation}
  \label{eq:III.11}
  \ND_{(a)}\doteqdot\int d^3\vec{r}\;\bar{\psi}_a(\vec{r})\gamma^0\psi(\vec{r})-1\equiv 0\ .
\end{equation}

The second constraint to be imposed refers to the Poisson equations
(\ref{eq:II.43a})-(\ref{eq:II.43d}). Multiplying through these equations by the
corresponding potentials and integrating by parts over whole 3-space (under use of Gau\ss'
integral theorem) yields the following \emph{Poisson identities}~\cite{MaSo,BMS}
\begin{subequations}
  \begin{align}
    \label{eq:III.12a}
    \NGe_{(1)} &\doteqdot \int d^3\vec{r}\;\left[\vec{\nabla}\,\iAn\sdot\vec{\nabla}\iiAn
      +4\pi\as\iAn\cdot\iikn\right] \equiv 0\\*
    \label{eq:III.12b}
    \NGe_{(2)} &\doteqdot \int d^3\vec{r}\;\left[\vec{\nabla}\,\iiAn\sdot\vec{\nabla}\iAn
      -4\pi\as\iiAn\cdot\ikn\right] \equiv 0\\*
    \label{eq:III.12c}
    \NGm_{(1)} &\doteqdot \int d^3\vec{r}\;\left[ \left(\vec{\nabla}\times\vec{A}_1(\vec{r})\right)
      \sdot \left(\vec{\nabla}\times\vec{A}_2(\vec{r})\right)
      +4\pi\as\vec{A}_1(\vec{r})\sdot\vec{k}_2(\vec{r})\right] \equiv 0\\*
    \label{eq:III.12d}
    \NGm_{(2)} &\doteqdot \int d^3\vec{r}\;\left[ \left(\vec{\nabla}\times\vec{A}_2(\vec{r})\right)
      \sdot \left(\vec{\nabla}\times\vec{A}_1(\vec{r})\right)
      -4\pi\as\vec{A}_2(\vec{r})\sdot\vec{k}_1(\vec{r})\right] \equiv 0\ .
  \end{align}
\end{subequations}
These two kinds of constraints (i.e.\ wave function normalization and Poisson identities)
must now be implemented into the naive variational principle~$(\delta\ET=0)$ by means of the
method of Lagrangean multipliers so that one arrives at the modified variational principle
(~$\leadsto$ \emph{principle of minimal energy}):
\begin{equation}
  \label{eq:III.13}
  \delta\tET=0
\end{equation}
with the modified energy functional~$\tET$ being defined as the sum of the original
functional~$\ET$ (\ref{eq:III.1a}) and of the constraints
(\ref{eq:III.11})-(\ref{eq:III.12d}):
\begin{equation}
  \label{eq:III.14}
  \begin{split}
  \tET = &\left(\Z_{(1)}^2\cdot\Mp c^2 + \Z_{(2)}^2\cdot\Me c^2 \right) +
  2\left(\Tkin_{(1)}+\Tkin_{(2)}\right) + \left(\heER - \hmER\right)\\*
  &+ \sum_{a=1}^2\left(\lD_{(a)}\cdot\ND_{(a)} + \lGe_{(a)}\cdot\NGe_{(a)} +
  \lGma\cdot\NGm_{(a)} \right)\ .    
  \end{split}
\end{equation}
Here the magnetic mass equivalents~$\Mmiii c^2$ occurring in the original functional~$\ET$
(\ref{eq:III.1a}), namely via the matter energies~$\ED_{(a)}$
(\ref{eq:III.3a})-(\ref{eq:III.3b}), have been eliminated in favour of the magnetic field
energy~$\hmER$ (\ref{eq:III.9b}) by reference to the magnetic Poisson identities
(\ref{eq:III.12c})-(\ref{eq:III.12d}) which read in physical terms
\begin{equation}
  \label{eq:III.15}
  \hmER = - \Mmi c^2 = - \Mmii c^2\ .
\end{equation}

But here it is now a nice exercise to convince oneself of the pleasant fact that \emph{both} the
relativistic mass eigenvalue equations (\ref{eq:II.45a})-(\ref{eq:II.45b}) \emph{and} the
Poisson equations (\ref{eq:II.43a})-(\ref{eq:II.43d}) actually turn out as the variational
equations of the functional (\ref{eq:III.14}), provided one fixes the Lagrangean
multipliers~$\lD_{(a)}$ and~$\lGema$ as follows:
\begin{subequations}
  \begin{align}
    \label{eq:III.16a}
    \lD_{(1)} &= M_1 c^2\\*
    \label{eq:III.16b}
    \lD_{(2)} &= -M_2 c^2\\*
    \label{eq:III.16c}
    \lGe_{(a)} &= - \lGma = - \frac{\hbar c}{4\pi\as}\\*
    &(a=1,2)\ .\notag
  \end{align}
\end{subequations}

Concerning the general form of our final result~$\tET$ (\ref{eq:III.14}) it is interesting
to observe that this functional is built up by two different kinds of contributions,
namely the physical terms (first line) and the constraint terms (second line). Thus,when
one wishes to resort to some trial configurations for the purpose of minimalizing the
functional~$\tET$, it is advantageous to work with those ansatz functions which obey both
types of constraints (\ref{eq:III.11}) and (\ref{eq:III.12a})-(\ref{eq:III.12d}). In this
case, the second line on the right-hand side of equation (\ref{eq:III.14}) is zero and the
functional~$\tET$ becomes built up exclusively by the physical terms (first line). Both
cases (i.e.\ with vanishing and non-vanishing constraint terms) have been tested in the
preceding papers~\cite{MaSo,BMS} and it has been found that the theoretical predictions are in
better coincidence with the experimental data when the constraint terms are made zero by
an appropiate choice of the trial functions (see ref.~\cite{MaSo}). But if one should prefer
(from some reason) a variational ansatz which implies non-vanishing constraint terms, then
the non-trivial contribution of the latter terms ensures that the energy prediction~$\tET$
is kept in the vicinity of the proper energy value see ref.~\cite{BMS}.  Subsequently we
will use trial configurations with vanishing constraint terms and will then study the
associated (albeit non-relativistic) energy predictions for positronium.

\begin{center}
  \emph{\textbf{2.\ Non-Relativistic Approximation }}
\end{center}

The non-relativistic two-body problem is perfectly understood by the conventional theory,
both classically and quantum-mechanically~\cite{Ka}; and therefore any new relativistic
theory must not be admitted to come into conflict with the known (non-relativistic)
results of the conventional theory. For the present situation this means that the
non-relativistic positronium spectrum, as predicted by RST, should agree with the
corresponding conventional predictions (at least up to the relativistic corrections). Thus
it becomes necessary to look for the non-relativistic approximation~$(\tEnT)$ of the
relativistic energy functional~$\tET$ (\ref{eq:III.14}), so that the non-relativistic
positronium spectrum can be estimated by means of appropriate (but non-relativistic) trial
functions.

The non-relativistic form~$\tEnT$ of the energy functional~$\tET$ is, however, easily
deduced by simply looking for the non-relativistic forms of the individual contributions.
Thus, omitting as usual the rest masses for the non-relativistic approach, the first two
bracket terms on the right-hand side of equation (\ref{eq:III.14}) yield the ordinary
kinetic energies~$\Ekin_{(a)}$ of the two particles
\begin{equation}
  \label{eq:III.17}
  \left(\Z_{(1)}^2\cdot\Mp c^2 + \Z_{(2)}^2\cdot\Me c^2 \right) +
  2\left(\Tkin_{(1)}+\Tkin_{(2)}\right) \Rightarrow \Ekin_{(1)} + \Ekin_{(2)} \ ,
\end{equation}
with the non-relativistic energies being given by~\cite{BeSo}
\begin{subequations}
  \begin{align}
    \label{eq:III.18a}
      \Ekin_{(1)} &= \frac{\hbar^2}{4\Mp}\int d^2\vec{r}\left[
      \left(\frac{\partial \iRp}{\partial r}\right)^2 + \frac{1}{r^2}
      \left(\frac{\partial \iRp}{\partial \vartheta}\right)^2\right]\\*
    \label{eq:III.18b}
    \Ekin_{(2)} &= \frac{\hbar^2}{4\Me}\int d^2\vec{r}\left[
      \left(\frac{\partial \iiSp}{\partial r}\right)^2 + \frac{1}{r^2}
      \left(\frac{\partial \iiSp}{\partial \vartheta}\right)^2\right]\\*
      &\hspace{3cm}(d^2\vec{r} \doteqdot rdrd\vartheta)\ .\notag
   \end{align}
\end{subequations}

Next, neglecting the magnetic interactions for the present non-relativistic limit, the
general form of the electrostatic field energy~$\heER$ (\ref{eq:III.9a}) remains the same
for the non-relativistic limit; but the electric counterpart of the magnetic Poisson
identities, i.e.
\begin{equation}
  \label{eq:III.19}
  \heER = \Mei c^2 = \Meii c^2\ ,
\end{equation}
receives some change concerning the electric mass equivalents~$\Meiii c^2$
(\ref{eq:III.7a})-(\ref{eq:III.7b}). The point here is that the original relativistic form
of the Dirac densities~$\akn$
\begin{equation}
  \label{eq:III.20}
  \akn = \frac{\aRp^2+\aRm^2+\aSp^2+\aSm^2}{4\pi r\sin\vartheta}
\end{equation}
becomes truncated to the first contribution~$(\sim\aRp^2)$ as shown by equations
(\ref{eq:II.54a})-(\ref{eq:II.54b}). Indeed multiplying through the non-relativistic
Poisson equations (\ref{eq:II.55a})-(\ref{eq:II.55b}) and integrating by use of Gau\ss'
theorem yields the non-relativistic form of the electric Poisson identities
(\ref{eq:III.12a})-(\ref{eq:III.12b}) as
\begin{subequations}
  \begin{align}
    \label{eq:III.21a}
    \NGe_{(1)} &\Rightarrow \NGn_{(1)} \doteqdot \int d^3\vec{r}\;\left( \vec{\nabla}\, \iAn\sdot\vec{\nabla}\,
      \iiAn + \as\frac{\iAn\cdot\iiSp(\vec{r})^2 }{r\sin\vartheta} \right) \equiv 0\\*
    \label{eq:III.21b}
    \NGe_{(2)} &\Rightarrow \NGn_{(2)} \doteqdot \int d^3\vec{r}\;\left( \vec{\nabla}\, \iiAn\sdot\vec{\nabla}\,
      \iAn - \as\frac{\iiAn\cdot\iRp(\vec{r})^2}{r\sin\vartheta} \right) \equiv 0\ .
  \end{align}
\end{subequations}

Thus the final form of the desired non-relativistic energy functional~$\tEnT$ is the
following
\begin{equation}
  \label{eq:III.22}
  \begin{split}
  \tEnT = \Ekin_{(1)} + \Ekin_{(2)} + \heER +
  \sum_{a=1}^2 \left(\lS_{(a)}\cdot\NDn_{(a)} + \lGe_{(a)}\cdot\NGn_{(a)} \right) \ .\\
  \end{split}
\end{equation}
Here the non-relativistic kinetic energies~$\Ekin_{(a)}$ have been specified by equations
(\ref{eq:III.18a})-(\ref{eq:III.18b}); the non-relativistic electric field energy~$\heER$
preserves its relativistic form (\ref{eq:III.9a}) but is now related to the electric mass
equivalents~$\Meiii c^2$  (\ref{eq:III.7a})-(\ref{eq:III.7b}) in their non-relativistic form, i.e.
\begin{subequations}
  \begin{align}
    \label{eq:III.23a}
    \Mei c^2 &\Rightarrow \hbar c\int d^3\vec{r}\;\iiAn\cdot\frac{\iRp^2(\vec{r})}{4\pi r\sin\vartheta}\\*
    \label{eq:III.23b}
    \Meii c^2 &\Rightarrow -\hbar c\int d^3\vec{r}\;\iAn\cdot\frac{\iiSp^2(\vec{r})}{4\pi
      r\sin\vartheta}\ ,
  \end{align}
\end{subequations}
and the electric Poisson constraints~$\NGe_{(a)}$ have to be used in their non-relativistic
form~$\NGn_{(a)}$ (\ref{eq:III.21a})-(\ref{eq:III.21b}). Finally, the non-relativistic form of the
normalization conditions (\ref{eq:III.11}) is now
\begin{subequations}
  \begin{align}
    \label{eq:III.24a}
    \ND_{(1)} &\Rightarrow \NDn_{(1)} \doteqdot\int d^3\vec{r}\;
    \frac{\iRp^2(\vec{r})}{4\pi r\sin\vartheta} - 1 = 0\\*
    \label{eq:III.24b}
    \ND_{(2)} &\Rightarrow \NDn_{(2)} \doteqdot\int d^3\vec{r}\;
    \frac{\iiSp^2(\vec{r})}{4\pi r\sin\vartheta} - 1 = 0\ ,
  \end{align}
\end{subequations}
with the relativistic Lagrangean multipliers~$\lD_{(a)}$ adopting their non-relativistic
form~$\lS_{(a)}$ as the Schr\"odinger energy eigenvalues~$\ES_{(a)}$
(\ref{eq:II.52})-(\ref{eq:II.53})
\begin{subequations}
  \begin{align}
    \label{eq:III.25a}
    \lD_{(1)} &= M_1 c^2 \Rightarrow \lS_{(1)} = -\ES_{(1)} \doteqdot \Mp c^2 + M_1 c^2\\*
    \label{eq:III.25b}
    \lD_{(2)} &= -M_2 c^2 \Rightarrow \lS_{(2)} = -\ES_{(2)} \doteqdot \Me c^2 - M_2 c^2\ .
  \end{align}
\end{subequations}

But with all these arrangements it becomes now a straightforward matter to convince
oneself that the non-relativistic eigenvalue equations (\ref{eq:II.52})-(\ref{eq:II.53})
together with the non-relativistic Poisson equations (\ref{eq:II.55a})-(\ref{eq:II.55b})
actually turn out as the variational equations~$(\delta\tEnT=0)$ due to the
non-relativistic energy functional~$\tEnT$! However despite of this pleasant result, we
will not use this functional~$\tEnT$ (\ref{eq:III.22}) in its present form for the
subsequent calculation of the non-relativistic positronium spectrum but will first resort
to a further simplification.

\begin{center}
  \emph{\textbf{3.\ Spherically Symmetric Approximation}}
\end{center}

The wave amplitudes~$\aRpm,\aSpm$ (\ref{eq:II.47a})-(\ref{eq:II.47b}) are assumed to
depend not only upon the radial variable~$(r)$ but also upon the angle~$\vartheta$ of the
spherical polar coordinates. This angular dependence must be taken into account also for
the positronium groundstate (as well as for the excited states of the same symmetry)
despite the fact that the groundstate is mostly the state of highest symmetry, i.e.\ the
spherical symmetry for the present situation. It is true, the non-relativistic eigenvalue
equations (\ref{eq:II.52})-(\ref{eq:II.53}) would admit spherically symmetric
solutions~$\iRp(r),\iiSp(r)$ provided the electric potentials~$\aAr\rt$ are also
symmetric:~$\aAr\rt\Rightarrow\aAe(r)$; but even for SO(3) symmetric amplitudes~$\iRp(r)$
and~$\iiSp(r)$ the potentials~$\aAn$ as solutions of the non-relativistic Poisson
equations (\ref{eq:II.55a})-(\ref{eq:II.55b}) can then obviously not be spherically
symmetric. But because of these anisotropic potentials, the solutions~$\iRp$ and~$\iiSp$
of the eigenvalue equations will then also be found to miss the spherical symmetry and this
contradicts the original assumption for the wave amplitudes. This anisotropic feature of
the RST groundstate solutions is in contrast to the SO(3) symmetry of the residual
internal Coulomb force problem in the conventional theory after the centre-of-mass motion
has been separated off~\cite{Me}. Thus, in order to compare the RST predictions for the
non-relativistic positronium states to the corresponding conventional predictions on the
same level of symmetry, one will restrict the RST trial functions to just this spherical
symmetry.  Furthermore, one assumes that both particles do occupy physically equivalent
states, i.e.\ the wave functions and potentials due to both particles must be identical.
This yields the following identifications~\cite{BMS}:
\begin{subequations}
  \begin{align}
    \label{eq:III.26a}
    \iRp\rt &\equiv \iiSp\rt\doteqdot\tilde{R}(r)\\*
    \label{eq:III.26b}
    {}^{(1)}\!k_0\rt &\equiv {}^{(2)}\!k_0\rt\doteqdot
    \frac{\btkn}{4\pi r\sin\vartheta} = \frac{\tilde{R}(r)^2}{4\pi r\sin\vartheta}\\*
    \label{eq:III.26c}
    {}^{(1)}\!A_0\rt&\equiv - {}^{(2)}\!A_0\rt \doteqdot\ppAn(r)\ .  
  \end{align}
\end{subequations}
Accordingly, the two non-relativistic eigenvalue equations  (\ref{eq:II.52}) and
(\ref{eq:II.53}) become contracted to a single one
\begin{equation}
  \label{eq:III.27}
  -\frac{\hbar^2}{2M}\left[\frac{1}{r}\cdot\frac{d}{dr}\left(r\cdot\frac{d\tilde{R}(r)}{dr}
    \right) \right] - \hbar c\; \ppAn(r) \cdot\tilde{R}(r) = \ES\cdot\tilde{R}(r)\ .
\end{equation}

On the other hand, the deduction of the spherically-symmetric form of the two associated
Poisson equations (\ref{eq:II.55a})-(\ref{eq:II.55b}) is not so evident and requires an
extra argument: For obtaining the SO(3) symmetric form of the non-relativistic energy
functional~$\tEnT$ (\ref{eq:III.22}) one observes first that by the present
identifications (\ref{eq:III.26a})-(\ref{eq:III.26b}) the kinetic energies~$\Ekin_{(a)}$
(\ref{eq:III.18a})-(\ref{eq:III.18b}) become identical
\begin{equation}
  \label{eq:III.28}
  \Ekin_{(1)} = \Ekin_{(2)}\doteqdot\Ekin=\pi\frac{\hbar^2}{4M}\cdot\int_0^\infty dr\,r
  \left(\frac{d\tilde{R}(r)}{dr}\right)^2\ .
\end{equation}
Next, the spherically symmetric form of the electric field energy~$\heER$
(\ref{eq:III.9a}) becomes by reference to the isotropic potential~$\ppAn(r)$
(\ref{eq:III.26c})
\begin{equation}
  \label{eq:III.29}
  \heER\Rightarrow-\frac{\hbar c}{\as}\int_0^\infty dr\,r^2\left(\frac{d\,\ppAn(r)}{dr}\right)^2\ ,
\end{equation}
and finally the non-relativistic normalization conditions
(\ref{eq:III.24a})-(\ref{eq:III.24b}) contract to a single one for the residual wave
amplitude~$\tilde{R}(r)$ (\ref{eq:III.26a}) 
\begin{equation}
  \label{eq:III.30}
  \NDn_{(1)}=\NDn_{(2)}\doteqdot\NDn=\frac{\pi}{2}\int_0^\infty dr\,r\tilde{R}(r)^2-1=0\ .
\end{equation}
Of course, both non-relativistic Poisson identities (\ref{eq:III.21a})-(\ref{eq:III.21b})
do also collapse to a single one:
\begin{equation}
  \label{eq:III.31}
  \begin{split}
    \NGn_{(1)}&=\NGn_{(2)}\doteqdot\NGn=4\pi\int_0^\infty dr\,r^2\left[-\left(\frac{d\,\ppAn(r)}{dr}
      \right)^2 + \frac{\pi}{2}\as\frac{ \ppAn(r) \cdot\tilde{R}(r)^2}{r} \right]\\*
    &=0\ .    
  \end{split}
\end{equation}
Thus  the spherically symmetric form of the non-relativistic functional~$\tEnT$
(\ref{eq:III.22}) is found to look as follows
\begin{equation}
  \label{eq:III.32}
  \begin{split}
  \tEnT\Rightarrow&\frac{\pi}{2}\frac{\hbar^2}{M}\int_0^\infty dr\,r
  \left(\frac{d\tilde{R}(r)}{dr} \right)^2 - \frac{\hbar c}{\as}\int_0^\infty dr\,r^2
  \left(\frac{d\,\ppAn(r)}{dr} \right)^2 \\*
  &+ 2\lS\cdot\NDn + 2\lGe\cdot\NGn\ ,
  \end{split}
\end{equation}
where the Lagrangean multipliers for the wave function normalization
(\ref{eq:III.24a})-(\ref{eq:III.24b}) become identical, too, and equal then the common
non-relativistic energy eigenvalue~$\ES$
\begin{equation}
  \label{eq:III.33}
  \lS_{(1)} = \lS_{(2)} \doteqdot\lS=-\ES
\end{equation}
(the multipliers (\ref{eq:III.16c}) for the electric Poisson constraints  are identical in
any case $\left(\lGe_{(1)})=\lGe_{(2)}\doteqdot\lGe\right)$).

Now it is again an easy exercise to convince oneself of the fact that the SO(3) invariant
functional~$\tEnT$ (\ref{eq:III.32}) actually is stationary upon the
solutions~$\tilde{R}(r)$ of the SO(3) invariant eigenvalue equation (\ref{eq:III.27}) as
required. But the point here is that the desired Poisson equation for the rotationally
invariant potentials~$\ppAn(r)$ is obtained now as the corresponding variational equation of
the present energy functional (\ref{eq:III.32}):
\begin{equation}
  \label{eq:III.34}
  \frac{1}{r^2}\frac{d}{dr}\left(r^2\cdot\frac{d\,\ppAn(r)}{dr} \right) = -\frac{\pi}{2}\as
  \cdot \frac{\tilde{R}(r)^2}{r}\ .
\end{equation}
This completes the non-relativistic, spherically symmetric RST picture of the positronium
system. It consists of the eigenvalue equation for the common wave
amplitude~$\tilde{R}(r)$ (\ref{eq:III.27}) and the Poisson equation (\ref{eq:III.34}),
with the energy functional given by (\ref{eq:III.32}). The pleasant effect with this
result is now that the energy functional~$\tEnT$ (\ref{eq:III.32}) is stationary upon the
solutions of the coupled matter-gauge field system (\ref{eq:III.27}) and
(\ref{eq:III.34}), so that one can now guess appropriate trial functions for the wave
amplitude~$\tilde{R}(r)$ with ansatz parameters~$b_k$ in order to extremalize the energy
functional~$\tEnT$ (\ref{eq:III.32}) just with respect to these trial parameters~$b_k$.

\begin{center}
  \emph{\textbf{4.\ Poisson Identities}}
\end{center}

As mentioned above, the accuracy of the RST predictions for the total energy~$\ET$ become
improved when one does not resort to two \emph{rather unrelated} ans\"atze for the wave
amplitude~$\tilde{R}(r)$ and for the electric gauge potential~$\ppAn(r)$ but when one
observes the link between both ans\"atze which is established by the electric
\emph{Poisson identity}~$\NGn$ (\ref{eq:III.31}), or by (\ref{eq:III.19}), resp. Actually
such an improvement can be attained by choosing some trial ansatz for the wave
amplitude~$\tilde{R}(r)$ (as realistic as possible) and associating to it the
corresponding solution~$\ppAn(r)$ of the Poisson equation (\ref{eq:III.34}). Subsequently, we
will elaborate this point in great detail by means of a special trial
function~$\tilde{R}(r)$.

Here, it suggests itself to try an exponential~$(\sim\exp(-r/r_*))$ times some
other function~($P(r)$, say), i.e.\ our general trial ansatz will look as follows
\begin{equation}
  \label{eq:III.35}
  \tilde{R}(r) = \sqrt{\frac{8}{\pi r_*^2}}P(y)\exp(-\beta y)\ ,
\end{equation}
where~$r_*$ is some constant length parameter which mainly serves to obey the
(non-relativistic) normalization condition (\ref{eq:III.30}). The radial variable~$r$ of
the spherical polar coordinates is rescaled to the dimensionless variable~$y$
\begin{gather}
  \label{eq:III.36}
  y\doteqdot\frac{r}{\aB}\\*
  \Big(\aB=\frac{\hbar^2}{Me^2}\ldots\text{Bohr radius} \Big)\ ,\notag
\end{gather}
and the decay constant~$\beta$ is to be considered as one of the ansatz
parameters~$b_k$. The other parameters are thought to be contained in the ansatz
function~$P(y)$ which is required to not tend to infinity faster than~$\exp(\beta y)$
for~$y\to\infty$, so that the whole trial function~$\tilde{R}(r)$ safely tends to zero at
spatial infinity~$(y\to\infty)$. But clearly before one looks now for the associated
potential~$\ppAn(r)$ according to the Poisson equation, one will first satisfy the
normalization condition (\ref{eq:III.30}) which adopts the following shape
\begin{equation}
  \label{eq:III.37}
  \int_0^\infty dy\,y Q(y) e^{-2\beta y} = \left(\frac{y_*}{2}\right)^2
\end{equation}
where the constant~$y_*$ is related to the original normalization parameter~$r_*$
(\ref{eq:III.35}) through
\begin{equation}
  \label{eq:III.38}
  y_*\doteqdot\frac{r_*}{\aB}
\end{equation}
and the function~$Q(y)$ is the square of~$P(y)$, i.e.
\begin{equation}
  \label{eq:III.39}
  Q(y)\doteqdot P(y)^2\ .
\end{equation}

After the general form of the trial wave amplitude has thus been fixed, one next wishes to
specify also the general form of the associated solutions~$\ppAn(r)$ of the Poisson equation
(\ref{eq:III.34}). For this purpose one first writes down the formal solution of this
equation for the field strength~$\ppF$ as
\begin{equation}
  \label{eq:III.40}
  \ppF \doteqdot \frac{d\,\ppAn(r)}{dr} = -\frac{\as}{r^2}\left(1-\frac{\pi}{2}
  \int_r^\infty dr\,r\tilde{R}(r)^2\right)\ .
\end{equation}
Here one demands that the electric field strength (i.e.\ the gradient field of the
potential~$\pAn$) asymptotically adopts the usual Coulomb form due to one charge unit:
\begin{equation}
  \label{eq:III.41}
  \lim_{r\to\infty}\ppF = -\frac{\as}{r^2}\ .
\end{equation}
Clearly, this is easily seen to be a consequence of the normalization condition
(\ref{eq:III.30}). It is more convenient to recast the field strength~$\ppF$ to its
dimensionless form
\begin{equation}
  \label{eq:III.42}
  \ppF = -\frac{\as}{\aB^2}\cdot\frac{1}{y^2}\left[1-\left(1+\tilde{f}(y)\right)e^{-2\beta
    y} \right]\ ,
\end{equation}
where the dimensionless function~$\tilde{f}(y)$ has evidently been introduced through
\begin{equation}
  \label{eq:III.43}
  \tilde{f}(y) = -1 + e^{2\beta y}\cdot\frac{\pi}{2}\int_r^\infty dr'\,r'\tilde{R}(r')^2\ .
\end{equation}
Observe here that, by virtue of just the normalization condition (\ref{eq:III.30}), the
function~$\tilde{f}(y)$ vanishes at the origin~$(y=0)$ so that the electric field
strength~$\ppF$ (\ref{eq:III.42}) may adopt there a finite value (for an example of this
see ref.~\cite{BMS}). One may therefore put
\begin{equation}
  \label{eq:III.44}
  \tilde{f}(y) = 2\beta y + \tilde{G}(y)
\end{equation}
where~$\tilde{G}(y)$ must then tend to zero quadratically at the origin, i.e.
\begin{gather}
  \label{eq:III.45}
  \tilde{G}(y) = \tilde{g}_2\cdot y^2 + \mathrm{o}(y^3)\\*
  \big(\tilde{g}_2 = \text{const.} \big)\ .\notag
\end{gather}
Namely, by this arrangement the value of the field strength~$\ppF$ (\ref{eq:III.40}) at
the origin~$(y=0)$ is then found as
\begin{equation}
  \label{eq:III.46}
  \ppFn = -\frac{\as}{\aB^2}\left(2\beta^2-\tilde{g}_2\right)\ ,
\end{equation}
whereas its general form is found to look as follows:
\begin{equation}
  \label{eq:III.47}
  \ppF = -\frac{\as}{\aB^2}\cdot\frac{1}{y^2}\left(\hat{F}(y)-\tilde{G}(y)\cdot e^{-2\beta y}
  \right)\ ,
\end{equation}
with the dimensionless function~$\hat{F}(y)$ being given by
\begin{equation}
  \label{eq:III.48}
  \hat{F}(y) = 1 - \left(1+2\beta y \right)e^{-2\beta y}\ .
\end{equation}

The field strength~$\ppF$ (\ref{eq:III.47}) already displays its final form to be used
subsequently for the computation of the field energy~$\heER$ (\ref{eq:III.29}), and
therefore it is advantageous to link the function~$\tilde{G}(y)$ directly to the ansatz
function~$Q(y)$ (\ref{eq:III.39}). This, however, is easily achieved by substituting the
field strength~$\ppF$ (\ref{eq:III.47}) back into the Poisson equation (\ref{eq:III.34})
in order to deduce the desired differential equation for~$\tilde{G}(y)$ as
\begin{equation}
  \label{eq:III.49}
  -\frac{d\tilde{G}(y)}{dy}+2\beta\tilde{G}(y) = y\cdot\tilde{Q}(y)
\end{equation}
where the function~$\tilde{Q}(y)$ is a slight modification of the original~$Q(y)$
(\ref{eq:III.39}):
\begin{equation}
  \label{eq:III.50}
  \tilde{Q}(y)\doteqdot\left(\frac{2}{y_*}\right)^2 Q(y) - \left(2\beta\right)^2\ .
\end{equation}
This new function~$\tilde{Q}(y)$ obeys the integral relation
\begin{equation}
  \label{eq:III.51}
  \int_0^\infty dy\,y\tilde{Q}(y)e^{-2\beta y}=0\ ,
\end{equation}
which of course is an immediate consequence of the former normalization condition
(\ref{eq:III.37}). For the subsequent demonstrations with appropriate ansatz
functions~$\tilde{Q}(y)$, the associated constituent~$\tilde{G}(y)$ of the field
strength~$\ppF$ (\ref{eq:III.47}) is then easily calculated from the differential equation
(\ref{eq:III.49}) as
\begin{equation}
  \label{eq:III.52}
  \tilde{G}(y) = e^{2\beta y}\cdot\int_y^\infty dy'\,y'\tilde{Q}(y')e^{-2\beta y'}\ .
\end{equation}

However with respect to the Poisson identity, the most important feature of the field
strength~$\ppF$ (\ref{eq:III.47}) refers to its sum structure which implies the splitting
of the field energy~$\heER$ (\ref{eq:III.29}) into three contributions
\begin{equation}
  \label{eq:III.53}
  \begin{split}
    \heER &= -\frac{\hbar c}{\as}\int_0^\infty dr\,r^2\left(\ppF\right)^2\\*
    &=-\frac{e^2}{\aB}\int_0^\infty dy\left(\frac{\hat{F}(y)}{y} \right)^2 +
    2\frac{e^2}{\aB}\int_0^\infty\frac{dy}{y^2}\;e^{-2\beta y}\hat{F}(y)\tilde{G}(y)\\*
    &- \frac{e^2}{\aB}\int_0^\infty dy\left(\frac{e^{-2\beta y}\tilde{G}(y)}{y} \right)^2\ .
  \end{split}
\end{equation}
The first contribution (due to~$\hat{F}(y)$) essentially represents the field energy of
the groundstate, whereas the second (quadratic) term and the third (quartic) term
describe the influence of the excited states (see below). Observe also that all three
integrals in (\ref{eq:III.53}) assume finite values, namely since the field strength~$\ppF$
(\ref{eq:III.47}) remains finite at the origin~$(y=0)$, cf.~(\ref{eq:III.46}).

With the field energy~$\heER$ being analyzed now in sufficient detail, a similar
investigation of the electric mass equivalents~$\Meiii c^2$ is necessary in order to
elucidate the Poisson identities (\ref{eq:III.19}) more thoroughly. First observe here
that, by virtue of the groundstate identifications (\ref{eq:III.26a})-(\ref{eq:III.26c}),
both electric mass equivalents~$\Meiii c^2$  (\ref{eq:III.23a})-(\ref{eq:III.23b}) become
identical
\begin{equation}
  \label{eq:III.54}
  \Mei c^2 = \Meii c^2 \doteqdot M^\textrm{(e)} c^2 = -\frac{\pi}{2}\hbar c
  \int_0^\infty dr\,r\,\ppAn(r)\, \btknn\ .
\end{equation}
Here, the common reduced Dirac density~$\btknn$ is given by the non-relativistic
approximations together with the groundstate identifications as
\begin{equation}
  \label{eq:III.55}
  \btknn \equiv \tilde{R}(r)^2\ ,
\end{equation}
and furthermore the electric interaction potential~$\ppAn(r)$ is decomposed, analogously to
its field strength~$\ppF$ (\ref{eq:III.47}), into the groundstate
contribution~$\hat{A}(y)$ and the excitation part~$\tilde{A}(y)$ as follows:
\begin{equation}
  \label{eq:III.56}
  \ppAn(r) = \frac{\as}{\aB}\cdot\frac{1}{y}\left(\hat{A}(y)-\tilde{A}(y)\exp[-2\beta y]
  \right)\ .
\end{equation}
Both constituents~$\hat{A}(y)$ and~$\tilde{A}(y)$ must of course be determined by
identifying the derivative of the potential~$\ppAn(r)$ with its field strength~$\ppF$
(\ref{eq:III.47}) which then yields the following two differential equations:
\begin{subequations}
  \begin{align}
    \label{eq:III.57a}
    \hat{A}(y) - y\frac{d\hat{A}(y)}{dy} &= \hat{F}(y)\\*
    \label{eq:III.57b}
    \left(1+2\beta y\right)\tilde{A}(y) - y\frac{d\tilde{A}(y)}{dy} &= \tilde{G}(y)\ .
  \end{align}
\end{subequations}

The first one (\ref{eq:III.57a}) of these equations can be immediately solved because the
right-hand side~$\hat{F}(y)$ is explicitly known, cf.~(\ref{eq:III.48}), which yields
\begin{equation}
  \label{eq:III.58}
  \hat{A}(y) = y\cdot\int_y^\infty dy'\,\frac{\hat{F}(y')}{y'{}^2}=1-e^{-2\beta y}\ .
\end{equation}
The second equation (\ref{eq:III.57b}) can of course not be solved explicitly since the
right-hand side~$\tilde{G}(y)$ will depend upon the selected trial ansatz,
cf.~(\ref{eq:III.52}); but for the practical purpose (see below) it is convenient to look
for the formal solution~$\tilde{A}(y)$
\begin{equation}
  \label{eq:III.59}
  \tilde{A}(y) = ye^{2\beta y}\cdot\int_y^\infty \frac{dy'}{y'{}^2} e^{-2\beta
    y'}\tilde{G}(y')\ .
\end{equation}
Thus the excitation part~$\tilde{A}(y)$ of the potential~$\ppAn(r)$ (\ref{eq:III.56}) is
non-singular at the origin~$(y=0)$ and actually becomes zero:~$\tilde{A}(0)=0$, according
to the behavior of the function~$\tilde{G}(y)$ (\ref{eq:III.45}). This, however, then
implies that the electric potential~$\ppAn(r)$ (\ref{eq:III.56}) itself is finite at the
origin, i.e.
\begin{equation}
  \label{eq:III.60}
  {{}^{[\textrm{p}]}\!A_0(0)} = 2\beta\cdot\frac{\as}{\aB}
\end{equation}
(for a sketch of this type of potential see ref.~\cite{BMS}). Amazingly enough this value at
the origin does depend solely upon the ansatz parameter~$\beta,$ but not upon the other
ansatz parameters~$b_k$ which are contained in the ansatz function~$P(y)$
(\ref{eq:III.35}). Obviously, a pleasant property of these Coulomb-like potentials is that
they yield a well-defined \emph{finite} field energy~$\heER$ (\ref{eq:III.53}), in
contrast to the situation with the exact Coulomb potential. Clearly, when the field
energy~$\heER$ would be infinite, such relations like the Poisson identities
(\ref{eq:III.19}) would be meaningless!

With the general shape of both the electric potentials~$\hat{A}(y),\tilde{A}(y)$ and the
trial function~$\tilde{Q}(y)$ being fixed now through the equations (\ref{eq:III.50}) and
(\ref{eq:III.58})-(\ref{eq:III.59}), the corresponding general form of the electric mass
equivalent~$\Mee c^2$ (\ref{eq:III.54}) is found to look as follows:
\begin{equation}
  \label{eq:III.61}
  \begin{split}
    \Mee c^2 &= -\frac{e^2}{\aB}\cdot\left(2\beta\right)^2\int_0^\infty dy\,\hat{A}(y)\cdot
    e^{-2\beta y}\\*
    &-\frac{e^2}{\aB}\int_0^\infty dy\,\hat{A}(y)\tilde{Q}(y)\cdot e^{-2\beta y} +
    \frac{e^2}{\aB}\cdot \left(2\beta\right)^2\int_0^\infty dy\,\tilde{A}(y)\cdot
    e^{-4\beta y}\\*
    &+\frac{e^2}{\aB}\int_0^\infty dy\,\tilde{A}(y)\tilde{Q}(y)\cdot e^{-4\beta y}\ .
  \end{split}
\end{equation}

Now according to the Poisson identities, this result for the mass equivalent must be
exactly identical to the corresponding result for the field energy~$\heER$
(\ref{eq:III.53})! But here it appears as a matter of course that the required identity of
both results must already apply to the \emph{individual} contributions themselves. Thus,
considering first both groundstate contributions, one expects that the following identity
holds:
\begin{equation}
  \label{eq:III.62}
  \int_0^\infty dy\,\left(\frac{\hat{F}(y)}{y} \right)^2 =
  \left(2\beta\right)^2\int_0^\infty dy\,\hat{A}(y) e^{-2\beta y}\ .
\end{equation}
However, the validity of this relation is easily checked by simply substituting for the
functions~$\hat{F}(y)$ and~$\hat{A}(y)$ those results obtained through equations
(\ref{eq:III.48}) and (\ref{eq:III.58}) which yields
\begin{equation}
  \label{eq:III.63}
  \int_0^\infty dy\,\left(\frac{\hat{F}(y)}{y}\right)^2 = \beta
\end{equation}

Next, the Poisson identity demands the identity of the quadratic terms in
(\ref{eq:III.53}) and  (\ref{eq:III.61}), i.e.
\begin{equation}
  \label{eq:III.64}
  2\cdot\int_0^\infty \frac{dy}{y^2}\, e^{-2\beta y}\cdot\hat{F}(y)\tilde{G}(y) =
  \left(2\beta\right)^2\int_0^\infty dy\,\tilde{A}(y)\cdot e^{-4\beta y} -
  \int_0^\infty dy\,\hat{A}(y)\tilde{Q}(y)\cdot e^{-2\beta y}
\end{equation}
But here one substitutes for the function~$\hat{F}(y)$ the derivative of~$\hat{A}(y)$ from
equation (\ref{eq:III.57a}) and obtains through integrating by parts with regard of the
differential equation for~$\tilde{G}(y)$ (\ref{eq:III.49}) the following relation
\begin{equation}
  \label{eq:III.65}
  \int_0^\infty \frac{dy}{y^2}\,e^{-2\beta y}\cdot\hat{F}(y)\tilde{G}(y)  = -
  \int_0^\infty dy\,e^{-2\beta y}\cdot\hat{A}(y)\tilde{Q}(y)\ .
\end{equation}
This recasts the required identity for the quadratic terms (\ref{eq:III.64}) to the
following form
\begin{equation}
  \label{eq:III.66}
   \int_0^\infty dy\,e^{-2\beta y}\cdot\hat{A}(y)\tilde{Q}(y) = - \left(2\beta\right)^2
   \int_0^\infty dy\,\tilde{A}(y) e^{-4\beta y}
\end{equation}
whose correctness is easily proven by use again of the link of~$\tilde{A}(y)$
to~$\tilde{G}(y)$ (\ref{eq:III.57b}). Thus the quadratic terms of the field energy~$\heER$
(\ref{eq:III.53}) are actually identical to those of its mass equivalent~$\Mee c^2$,
(\ref{eq:III.61}).

Finally, it remains to prove also the identity of the quartic terms, i.e.
\begin{equation}
  \label{eq:III.67}
  \int_0^\infty dy\,\left(\frac{e^{-2\beta y}\cdot\tilde{G}(y)}{y} \right)^2 = -
  \int_0^\infty dy\, \tilde{A}(y)\tilde{Q}(y)\cdot e^{-4\beta y}\ .
\end{equation}
The correctness of this equation is easily realized by substituting for the modified
function~$\tilde{Q}(y)$ from equation (\ref{eq:III.49}) and integrating by parts with use
of the differential equation for~$\tilde{A}(y)$ (\ref{eq:III.57b}). This completes the
demonstration of the electric Poisson identity in terms of the auxiliary
functions~$\tilde{Q}(y),\tilde{G}(y)$ and~$\tilde{A}(y)$ which build up the interaction
potential~$\ppAn(r)$ (\ref{eq:III.56}) and its field strength~$\ppF$ (\ref{eq:III.47}).
Indeed for any trial function~$\tilde{R}(y)$, these three functions must be calculated
explicitly in order to obtain the field energy~$\heER$ (or its mass equivalent~$\Mee c^2$)
as part of the total energy~$\tEnT$ (\ref{eq:III.32}). Since the required trial
functions~$\tilde{R}(r)$ may possibly look very complicated, the computation of the total
energy~$\tEnT$ will also be somewhat intricate; and therefore the confidence in the
results may be supported by testing explicitly the present identities (\ref{eq:III.65},
(\ref{eq:III.66})) and (\ref{eq:III.67}). This procedure is exemplified now by a further
specialization of the trial functions~$\tilde{R}(r)$ (\ref{eq:III.35}).

\section{Hydrogen-like Wave Functions}
\indent

Surely, it will be very difficult (if not impossible) to find exact analytic solutions of
the present (non-relativistic) eigenvalue problem (\ref{eq:III.27}) plus
(\ref{eq:III.34}). But nevertheless it may be possible to get the corresponding energy
eigenvalue~$\tEnT$ very accurately if one succeeds to guess some trial
function~$\tilde{R}$ as realistic as possible. Here it was observed that for the
(non-relativistic) groundstate a simple exponential trial function~$\tilde{R}(r)$ is
sufficient in order to reproduce \emph{exactly} the groundstate energy~$E_0$ of the
conventional Schr\"odinger theory, see below and ref.~\cite{MaSo}. Incidentally such a simple
exponential function represents the \emph{exact} groundstate solution of the
\emph{conventional} theory which, by separating off the centre-of-mass motion, leads to
the ordinary Coulomb force problem for the relative motion of both particles. Therefore it
appears self-suggesting to select as the RST trial functions just that type of exact
solutions for the conventional Coulomb force problem, i.e.\ the \emph{hydrogen-like wave
  functions}. These are given by the product of some polynomial times an exponential
function. In this sense, we specialize now the hitherto undetermined ansatz
function~$P(y)$ (\ref{eq:III.35}) down to a polynomial of degree~$\NN$
\begin{equation}
  \label{eq:IV.1}
  P(y) \Rightarrow P_\NN(y) = \sum_{n=0}^\NN b_n y^n
\end{equation}
where the coefficients~$b_n (n=0,1,2\ldots \NN)$ work as the ansatz parameters (besides the
parameter~$\beta$) and the variable~$y$ is the rescaled radial distance~$r$
(\ref{eq:III.36}). Now, for inspecting more closely the groundstate and the excited
states, it is very instructive for the practical use of the RST principle of minimal energy
to first convince oneself of the validity of the Poisson identities in terms of the
polynomials~$P_\NN(y)$. 

Turning here first to the modified square~$\tilde{Q}(y)$ (\ref{eq:III.50}) of the
polynomial~$P_\NN(y)$, one gets a polynomial~$\tilde{Q}_{2\NN}(y)$ of degree~$2\NN$, i.e.
\begin{equation}
  \label{eq:IV.2}
  \tilde{Q}(y)\Rightarrow\tilde{Q}_{2\NN}(y) = \sum_{m,n=0}^\NN b'_m b'_n y^{m+n} -
  (2\beta)^2\ ,
\end{equation}
with the rescaled ansatz parameters~$b'_n$ being simply defined through
\begin{equation}
  \label{eq:IV.3}
  b'_n \doteqdot \frac{2}{y_*}\cdot b_n\ .
\end{equation}
Here, the constant~$y_*$ itself becomes a function of the ansatz parameters~$b_n$ just by
virtue of the normalization requirement (\ref{eq:III.37})
\begin{equation}
  \label{eq:IV.4}
  \left(\frac{y_*}{2}\right)^2 = \sum_{m,n=0}^\NN b_m
  b_n\frac{(m+n+1)!}{(2\beta)^{m+n+2}}\ ,
\end{equation}
where the first coefficient~$b_0$ may be adopted to be unity~$(b_0=1)$, without loss of
generality. Alternatively, the normalization condition (\ref{eq:IV.4}) may be recast also
to the following form
\begin{equation}
  \label{eq:IV.5}
  {b'_0}^2 - (2\beta)^2 = - \sum_{\stackrel{\scr n,m=0}{(n+m\ge 1)}}^\NN b'_m b'_n\cdot \frac{(m+n+1)!}{(2\beta)^{m+n}}
\end{equation}
which will be needed hereafter for the explicit calculation of the energy
eigenvalue~$\tEnT$. Observe also that, with the help of the latter form (\ref{eq:IV.5}) of
the normalization condition, the modified square~$\tilde{Q}_{2\NN}(y)$ (\ref{eq:IV.2}) can
be rewritten in the following form for~$\NN\ge 1$
\begin{subequations}
  \begin{align}
    \label{eq:IV.6a}
  \tilde{Q}_{2\NN}(y) &= \sum_{\stackrel{\scr n,m=0}{(n+m\ge 1)}}^\NN b'_m
  b'_n\cdot\tilde{q}_{m+n}(y)\\*
  \label{eq:IV.6b}
  \tilde{q}_k &\doteqdot y^k - \frac{(k+1)!}{(2\beta)^k}
  \end{align}
\end{subequations}
which is best suited to verify the integral relation (\ref{eq:III.51}), namely through the
somewhat trivial integral for any integer~$k$:
\begin{equation}
  \label{eq:IV.7}
  \int_0^\infty dy\,y\cdot\tilde{q}_k(y)e^{-2\beta y} = 0\ .
\end{equation}

With a consistent variational ansatz being now at hand, one can proceed to compute the
corresponding electrostatic objects in order to test the validity of the Poisson identity
just for that specific ansatz. First, turn to the function~$\tilde{G}(y)$ as a constituent
of the electric field strength~$\ppF$ (\ref{eq:III.47}) and find from the constructive
relation (\ref{eq:III.52}) for the corresponding polynomial of degree~$2\NN+1$:
\begin{equation}
  \label{eq:IV.8}
  \tilde{G}(y)\Rightarrow\tilde{G}_{2\NN+1}(y)=-(1+2\beta y)+\sum_{m,n=0}^\NN
  b'_mb'_n(m+n+1)!\cdot \tilde{g}_{mn}(y)
\end{equation}
with the functions~$\tilde{g}_{mn}(y)$ being defined through
\begin{equation}
  \label{eq:IV.9}
  \tilde{g}_{mn}(y)=\sum_{\nu=0}^{m+n+1}\frac{y^{m+n+1-\nu}}{(m+n+1-\nu)!(2\beta)^{\nu+1}}\ .
\end{equation}
But with the functions~$\tilde{G}_{2\NN+1}$ (\ref{eq:IV.8}) and~$\tilde{Q}_{2\NN}(y)$
(\ref{eq:IV.6a})-(\ref{eq:IV.6b}) being explicity known, one can check now the validity
of the first integral identity (\ref{eq:III.65}) and one actually finds its polynomial
realization in the following form:
\begin{equation}
  \label{eq:IV.10}
  \begin{split}
  \int_0^\infty\frac{dy}{y^2}e^{-2\beta y}\cdot\hat{F}(y)\tilde{G}_{2\NN+1}(y) 
  & = - \int_0^\infty dy\, e^{-2\beta y}\cdot\hat{A}(y)\tilde{Q}_{2\NN}(y) \\*
  &= \beta\cdot\hspace{-4mm}\sum_{\stackrel{\scr n,m=0}{(n+m\ge 1)}}^\NN p_mp_n(m+n)!\varkappa_{mn}\ .      
  \end{split}
\end{equation}
Here for the sake of simplicity, the original polynomial coefficients~$b'_m$
(\ref{eq:IV.3}) have been modified to their more convenient form~$p_m$ through
\begin{equation}
  \label{eq:IV.11}
  p_m\doteqdot\frac{b'_m}{(2\beta)^{m+1}}\ ,
\end{equation}
and furthermore the coefficients~$\varkappa_{mn}$ are given by 
\begin{equation}
  \label{eq:IV.12}
  \varkappa_{mn} = 1-m-n-\frac{1}{2^{m+n}}\ .
\end{equation}

Next, consider the auxiliary potential~$\tilde{A}(y)$ as part of the gauge field~$\ppAn(r)$
(\ref{eq:III.56}) which becomes a polynomial~$\tilde{A}_{2\NN}(y)$ of degree~$2\NN$ and whose
differential equation is then given by (\ref{eq:III.57b}) with the formal solution being
specified by equation (\ref{eq:III.59}), i.e.
\begin{equation}
  \label{eq:IV.13}
  \tilde{A}(y)\Rightarrow\tilde{A}_{2\NN}(y) = ye^{2\beta
    y}\cdot\int_{y'=y}^\infty\frac{dy'}{y'^2}e^{-2\beta y'}\cdot\tilde{G}_{2\NN+1}(y')\ .
\end{equation}
Substituting here the polynomial~$\tilde{G}_{2\NN+1}(y)$ from the result (\ref{eq:IV.8})
yields for the desired polynomial~$\tilde{A}_{2\NN}(y)$ the following form, simply by
straight-forward calculation:
\begin{equation}
  \label{eq:IV.14}
  \tilde{A}_{2\NN}(y) = \sum_{\stackrel{\scr n,m=0}{(n+m\ge 1)}}^\NN b'_mb'_n(m+n+1)!\,\tilde{a}_{mn}(y)\ .
\end{equation}
Here, the coefficient functions~$\tilde{a}_{mn}(y)$ are given by
\begin{subequations}
  \begin{align}
    \label{eq:IV.15a}
    \tilde{a}_{mn}(y) &=
    \sum_{\nu=0}^{m+n-1}\frac{\tilde{a}^{(\nu)}_{mn}(y)}{(m+n+1-\nu)!(2\beta)^{\nu+1}}\\*
    \label{eq:IV.15b}
    \tilde{a}^{(\nu)}_{mn}(y) &= \sum_{\mu=0}^{m+n-1-\nu} \frac{(m+n-1-\nu)!}{(m+n-1-\nu-\mu)!}\cdot
    \frac{y^{m+n-\nu-\mu}}{(2\beta)^{\mu+1}}\ .
  \end{align}
\end{subequations}
Thus, one can easily convince oneself of the fact that the polynomial~$\tilde{A}_{2\NN}(y)$
is of the order~$\tilde{A}_{2\NN}(y)=o(y)$ at the origin~$(y\to 0)$. This verifies
explicitly the conclusion (\ref{eq:III.60}) drawn from the general form of~$\tilde{A}(y)$
(\ref{eq:III.59}). And furthermore with the polynomial~$\tilde{A}_{2\NN}(y)$ being known, it
becomes a simple exercise to check also the validity of the second integral identity
(\ref{eq:III.66}) in its polynomial form:
\begin{equation}
  \label{eq:IV.16}
  \int_0^\infty dy\,e^{-2\beta y}\hat{A}(y)\tilde{Q}_{2\NN}(y) =
  -(2\beta)^2\int_0^\infty dy\,\tilde{A}_{2\NN}(y)e^{-4\beta y}\ .
\end{equation}
The explicit numerical realization of this identity can be deduced from the first
identity(\ref{eq:IV.10}).

Finally, the quartic identity (\ref{eq:III.67}) is to be specified down to the polynomial
ansatz which then yields by straight-forward calculation
\begin{equation}
  \label{eq:IV.17}
  \begin{split}
  &\int_0^\infty dy \left(\frac{e^{-2\beta y}\tilde{G}_{2\NN+1}(y)}{y}\right)^2 =
  -\int_0^\infty dy\tilde{A}_{2\NN}(y)\tilde{Q}_{2\NN}(y)e^{-4\beta y}\\*
  &=\beta\cdot\hspace{-4mm}\sum_{\stackrel{\scr n,m=0}{(n+m\ge 1)}}^\NN\sum_{\stackrel{\scr l,q=0}{(l+q\ge 1)}}^\NN
  p_mp_np_lp_q\cdot\frac{(m+n+1)!(l+q+1)!}{2^{m+n+l+q+2}}\cdot\tau_{mn,lq}    
  \end{split}
\end{equation}
where the coefficients~$\tau_{mn,lq}$ are given by
\begin{equation}
  \label{eq:IV.18}
  \tau_{mn,lq} = \sum_{\nu=0}^{m+n-1}\sum_{\mu=0}^{l+q-1}2^{2+\mu+\nu}
  \frac{(m+n+l+q-\nu-\mu)!}{(m+n+1-\nu)!(l+q+1-\mu)!}\ .
\end{equation}
Obviously, this is the quartic counterpart of the quadratic identities (\ref{eq:IV.10})
and (\ref{eq:IV.16}). 

But now that all constituents of the field energy~$\heER$ (\ref{eq:III.53}), or its mass
equivalent~$\Mee c^2$ (\ref{eq:III.61}), resp., are explicitly known for the polynomial
ansatz, the final result appears in the following form:
\begin{equation}
  \label{eq:IV.19}
  \heER = -\frac{e^2}{\aB}\beta\big(1+\VN \pN \big)
\end{equation}
with the electric \emph{potential function}~$\VN$ being given by
\begin{equation}
  \label{eq:IV.20}
  \begin{split}
  \VN\pN &= 2\cdot\hspace{-4mm}\sum_{\stackrel{\scr n,m=0}{(n+m\ge 1)}}^\NN
  p_mp_n\cdot(m+n)!\,\varkappa_{mn}\\*
 &+ \sum_{\stackrel{\scr n,m=0}{(n+m\ge 1)}}^\NN \sum_{\stackrel{\scr l,q=0}{(l+q\ge 1)}}^\NN
  p_mp_np_lp_q\cdot\frac{(m+n+1)!(l+q+1)!}{2^{m+n+l+q+2}}\tau_{mn,lq}    
  \end{split}
\end{equation}

Fortunately, the potential function emerges here as an ordinary (but multi-dimensional)
polynomial when written in terms of the modified ansatz parameters~$p_m$
(\ref{eq:IV.11}). This completes and verifies the identity of the field energy~$\heER$ and
its mass equivalent~$\Mee c^2$, which is one of the two physical constituents of the
energy functional~$\tEnT$ (\ref{eq:III.22}). And furthermore the kinematical constraints
are automatically obeyed, too, since \emph{both} the normalization condition (\ref{eq:III.30})
for the trial function~$\tilde{R}(r)$ is satisfied in the form (\ref{eq:III.51})
\emph{and} the Poisson constraint (\ref{eq:III.31}) is satisfied in form of the integral
identities (\ref{eq:III.62})-(\ref{eq:III.67})! Observe, however that the lowest-order
coefficient~$b'_0$ is to be considered as a function of the other ansatz
parameters~$\beta$ and~$b'_k (k=1\ldots \NN)$, see equation (\ref{eq:IV.5}) which reads in
terms of the modified ansatz parameters~$p_m$ (\ref{eq:IV.11}):
\begin{equation}
  \label{eq:IV.21}
  p_0^2+\sum_{\stackrel{\scr n,m=0}{(n+m\ge 1)}}^\NN p_mp_n\cdot(m+n+1)! \equiv
  \sum_{m,n=0}^\NN p_mp_n(m+n+1)! = 1\ .
\end{equation}
Evidently, this constraint defines a certain compact subspace of the configuration space
parameterized by the set of ansatz parameters $\{\beta;p_0,p_1,\ldots p_\NN \}$.

In order to make complete the non-relativistic energy functional~$\tEnT$
(\ref{eq:III.22}), it becomes necessary to specify also the kinetic energy~$\Ekin$ of both
particles in terms of the polynomial ansatz. The spherically symmetric approximation
of~$\Ekin$ has already been determined in equation (\ref{eq:III.28}); and if the
polynomial ansatz (\ref{eq:IV.1}) for the wave function~$\tilde{R}(r)$ (\ref{eq:III.35})
is substituted therein, one arrives at the following form of the kinetic energy:
\begin{equation}
  \label{eq:IV.22}
  2\Ekin = \frac{e^2}{\aB}\beta^2\left(1+\TN\pN \right)\ ,
\end{equation}
with the \emph{kinetic function}~$\TN$ being given by
\begin{equation}
  \label{eq:IV.23}
  \TN\pN = -4\cdot\hspace{-2mm}\sum_{\stackrel{\scr m,n=0}{(m+n)\ge 1}}^\NN p_m p_n\cdot m^2(m+n-1)!
\end{equation}
Thus the value of the energy functional~$\tEnT$ upon the set of hydrogen-like wave
functions is explicitly known so that one can go to look now for the stationary points of the
corresponding function~$\tEnT[\beta;p_k]$ in any subspace of order~$\NN$.

Observe here for this purpose that the non-relativistic energy functional (\ref{eq:III.22}) in
its spherically symmetric approximation (\ref{eq:III.32}) emerges in the following form:
\begin{equation}
  \label{eq:IV.24}
  \tEnT \{\NN\} = \frac{e^2}{\aB}\big(\beta^2(1+\TN)-\beta(1+\VN)\big)\ .
\end{equation}
This is a very pleasant result because it admits to eliminate the exponential ansatz
parameter~$\beta$ for the extremalization procedure. Recall that this procedure consists
in nothing else than putting to zero all the partial derivatives of the energy
function~$\tEnT$ (\ref{eq:IV.24}) with regard of the normalization constraint
(\ref{eq:IV.21}). Thus the extremalization with respect to~$\beta$ yields
\begin{equation}
  \label{eq:IV.25}
  \frac{\partial \tEnT}{\partial\beta}=\frac{e^2}{\aB}\left(2\beta(1+\TN)-(1+\VN)\right)=0
\end{equation}
which fixes the parameter~$\beta$ in terms of the other parameters~$p_0,p_1,\ldots
p_{\textrm{N}}$ through
\begin{equation}
  \label{eq:IV.26}
  \beta=\frac{1+\VN}{2(1+\TN)}\ .
\end{equation}
Substituting this result for~$\beta$ back into the energy function (\ref{eq:IV.24})
recasts the latter to the form
\begin{equation}
  \label{eq:IV.27}
  \tEnT\{\NN\} = -\frac{e^2}{4\aB}\cdot\frac{(1+\VN)^2}{1+\TN}\ .
\end{equation}

From this result one concludes that the total energy~$\tEnT\{\NN\}$ is always negative
because the kinetic energy (\ref{eq:IV.22}) is always positive. Therefore one expects that
the RST principle of minimal energy will yield for any~$\NN$ the highest energy level
which is admitted by the applied variational ansatz. Whether an analogous conclusion does
hold also with respect to a lowest-possible energy level depends upon the existence of a
lower bound of the energy functional, but this is suggested by the subsequent numerical
calculations.  Observe also that the exponential parameter~$\beta$ does not enter the
normalization constraint (\ref{eq:IV.21}); and therefore the extremalization procedure is
reduced to merely an $\NN$-dimensional problem over just that compact submanifold which is
specified by the normalization constraint (\ref{eq:IV.21}).

\begin{center}
  \emph{\textbf{1.\ Zero-Order Approximation~$(\NN=0)$}}
\end{center}

The zero-order configuration~$(\NN=0)$ naturally represents the lowest-order approximation
to the positronium groundstate. Since this has already been discussed in two preceding
papers~\cite{MaSo,BMS}, it may be sufficient here to briefly exemplify the working of the
\emph{principle of minimal energy} in the simplest case. The point with this zero-order
approximation is that it sets the overall geometric background for the excited states and
yields the \emph{exact} groundstate energy of the conventional theory
(\ref{eq:I.3}). Therefore it is especially interesting to see whether and to what extent
this lowest-order result becomes modified subsequently by the higher-order
approximations~$(\NN\ge 1)$.

The polynomial~$\PN(y)$ (\ref{eq:IV.1}) of zero order is
\begin{equation}
  \label{eq:IV.28}
  P_0(y) = b_0\ (=\text{const.})
  \end{equation}
and thus the corresponding trial wave function~$\tilde{R}(r)$ (\ref{eq:III.35}) looks as
follows 
\begin{equation}
  \label{eq:IV.29}
  \begin{split}
  \tilde{R}(r)\Rightarrow \ntR(r) = \sqrt{\frac{8}{\pi r_*^2}}\,b_0\exp(-\beta y) &=
  \sqrt{\frac{2}{\pi\aB^2}}\, b'_0\exp(-\beta y)\\*
  &=\sqrt{\frac{2}{\pi\aB^2}}\cdot 2\beta p_0 e^{-\beta y}\ .
  \end{split}
\end{equation}
The normalization condition (\ref{eq:IV.5}) then identifies both ansatz parameters~$b'_0$
and~$\beta$, i.e.
\begin{equation}
  \label{eq:IV.30}
  b'_0 = 2\beta \Leftrightarrow p_0=1\ ,
\end{equation}
so that the normalization constraint (\ref{eq:III.30}) is actually obeyed:
\begin{equation}
  \label{eq:IV.31}
  \frac{\pi}{2}\int_0^\infty dr\,r\left(\ntR(r) \right)^2=(2\beta)^2\int_0^\infty dy\,y
  e^{-(2\beta)y} = 1\ .
\end{equation}

Furthermore, the electric field strength~$\ppF$ (\ref{eq:III.47}) becomes
\begin{equation}
  \label{eq:IV.32}
  \ppF\Rightarrow \nF(r) = -\frac{\as}{\aB^2}\cdot\frac{1}{y^2}\cdot\hat{F}(y)
\end{equation}
with the function~$\hat{F}(y)$ being given by equation (\ref{eq:III.48}). The reason for
this simple result is that the field strength constituent~$\tilde{G}(y)$ (\ref{eq:III.52})
vanishes since the function~$\tilde{Q}_{2N}(y)$ (\ref{eq:IV.2}) is also zero for~$\NN=0$:
\begin{equation}
  \label{eq:IV.33}
  \tilde{Q}_{2N}(y)\Rightarrow\tilde{Q}_0 = b'_0{}^2-(2\beta)^2 = 0\ .
\end{equation}
But clearly, if the electric field strength adopts such a simple form, the corresponding
potential~$\ppAn(r)$ (\ref{eq:III.56}) must be expected to be of a comparably simple
shape. However, one obviously has
\begin{equation}
  \label{eq:IV.34}
  \ppAn(r)\Rightarrow \nAn(r) = \frac{\as}{\aB}\cdot\frac{\hat{A}(y)}{y}
\end{equation}
with the function~$\hat{A}(y)$ being given by equation (\ref{eq:III.58}). The reason for
the simplicity of the groundstate result is again that the constituent~$\tilde{A}(y)$
(\ref{eq:III.59}) of the electric potential~$\ppAn(r)$ (\ref{eq:III.56}) is zero for the
groundstate ($\tilde{A}_{2\NN}(y)\Rightarrow\tilde{A}_0(y)$, cf.\ (\ref{eq:IV.13})). Observe
here that both the groundstate potential~$\nAn(r)$ (\ref{eq:IV.34}) and its field
strength~$\nF(r)$ (\ref{eq:IV.32}) are finite at the origin~$(y=0)$
\begin{subequations}
  \begin{align}
    \label{eq:IV.35a}
    \nAn\big|_{y=0} &= 2\beta\cdot\frac{\as}{\aB}\\*
    \label{eq:IV.35b}
    \nF\big|_{y=0} &= -2\beta^2\cdot\frac{\as}{\aB^2}\ ,
  \end{align}
\end{subequations}
which verifies the more general conclusion for~$\ppFn$ (\ref{eq:III.46}) and~$\ppAn(0)$
(\ref{eq:III.60}). For a sketch of this ``Coulomb-like'' type of potentials, see
ref.~\cite{BMS}. But clearly for the groundstate situation with vanishing
functions~$\tilde{Q}_0(y)$ (\ref{eq:IV.2}) and~$\tilde{G}_1(y)$ (\ref{eq:IV.8}), the
quadratic integral identity (\ref{eq:IV.10}) is satisfied trivially, as well as the
identity (\ref{eq:IV.16}) and the quartic identity (\ref{eq:IV.17}).

Naturally for such a simple field configuration, the corresponding total
energy~$\tEnT\{0\}$ (\ref{eq:IV.27}) must be expected to adopt a very simple form, too:
\begin{equation}
  \label{eq:IV.36}
  \tEnT\{0\} = -\frac{e^2}{4\aB}\cdot\frac{(1+V_0)^2}{1+T_0}=-\frac{e^2}{4\aB}\ ,
\end{equation}
since both the potential function~$\VN$ (\ref{eq:IV.20}) and the kinetic function~$\TN$
(\ref{eq:IV.23}) become zero for~$\NN=0$
\begin{equation}
  \label{eq:IV.37}
  V_0=T_0=0\ .
\end{equation}
The result (\ref{eq:IV.36}), however, is rather amazing because it just coincides with the
conventional groundstate prediction (\ref{eq:I.3}) for the principal quantum
number~$n_p=0$. Observe here, that the present RST result (\ref{eq:IV.36}) is an
\emph{approximation} due to the applied variational procedure, whereas the conventional
result (\ref{eq:I.3}) is usually considered to be \emph{exact within the framework of the
  conventional non-relativistic quantum mechanics}! (See appendix). Therefore it remains
to be discussed whether or not the \emph{true} (but hitherto unknown) non-relativistic
RST prediction does deviate from the (necessarily relativistic) experimental data~\cite{Ka}
to a smaller or larger extent in comparison to the deviations of the conventional
(non-relativistic) predictions, see below.

\begin{center}
  \emph{\textbf{2.\ First-Order Spectrum~$(\NN=1)$}}
\end{center}

Obviously, the general formalism in Sect.~III singles out the zero-order approximation as
an especially simple subcase. The reason for this is that the parameter~$\beta$ is of the
exponential type whereas all the other parameters~$b_n$ (\ref{eq:IV.1}) are of the
polynomial type. But on principle, the variational formalism works in an analogous manner
also for the higher-order approximations \mbox{$(\NN\ge 1)$}. Therefore it may be sufficient
here to collect, for the discussion of the first-order spectrum $(\NN=1)$, all the
prerequisites in a shortened procedure:

For~$\NN=1$, the kinetic function~$\TN$ (\ref{eq:IV.23}) is easily seen to adopt the
following simple shape
\begin{equation}
  \label{eq:IV.38}
  T_1(p_0,p_1) =- 4p_1(p_0+p_1)\ ,
\end{equation}
whereas the potential function~$\VN$ (\ref{eq:IV.20}) must necessarily look somewhat more
complicated: 
\begin{equation}
  \label{eq:IV.39}
  V_1(p_0,p_1) = -p_1(2p_0+5p_1-2p_0^2p_1-9p_0p_1^2-\frac{21}{2}p_1^3)\ .
\end{equation}
With these results, the first-order energy functional~$\tEnT\{1\}$ (\ref{eq:IV.27})
appears as
\begin{equation}
  \label{eq:IV.40}
  \tEnT\{1\} = -\frac{e^2}{4\aB}\cdot\frac{(1+V_1)^2}{1+T_1}
\end{equation}
and could now be extremalized under the normalization constraint (\ref{eq:IV.21}). The
latter is, for the present first-order situation~$(\NN=1)$, a simple quadratic form of two
variables~$(p_0,p_1)$, i.e.
\begin{equation}
  \label{eq:IV.41}
  p_0^2+4p_0p_1+6p_1^2=1\ .
\end{equation}

Such a simple extremalization problem over a compact configuration space can of course be
easily solved, but it is very instructive to look upon this procedure also from a somewhat
different viewpoint: Since the subspace defined by the constraint (\ref{eq:IV.41}) is
topologically equivalent to a circle~$S^1$, one introduces the angular variable~$\alpha\
(0\le\alpha\le2\pi)$ through
\begin{subequations}
  \begin{align}
    \label{eq:IV.42a}
    p_1 &= \frac{\sin\alpha}{\sqrt{2}}\\*
    \label{eq:IV.42b}
    p_0 &= \cos\alpha - \sqrt{2}\sin\alpha
  \end{align}
\end{subequations}
so that the normalization constraint (\ref{eq:IV.41}) is automatically
satisfied. Furthermore, the kinetic function~$T_1(p_0,p_1)$ (\ref{eq:IV.38}) and the
potential function~$V_1(p_0,p_1)$ (\ref{eq:IV.39}) become angular functions~$T_1(\alpha)$
and~$V_1(\alpha)$ through the parameterization (\ref{eq:IV.42a})-(\ref{eq:IV.42b}), i.e.
\begin{subequations}
  \begin{align}
    \label{eq:IV.43a}
    T_1(p_0,p_1) &\Rightarrow T_1(\alpha) =
    2\sin\alpha\left(\sin\alpha-\sqrt{2}\cos\alpha\right)\\*
    \label{eq:IV.43b}
    V_1(p_0,p_1)&\Rightarrow V_1(\alpha) = \sin\alpha\left(\frac{1}{2}\sin\alpha +
      \frac{\sqrt{2}}{4}\cos\alpha\sin^2\alpha - \sqrt{2}\cos\alpha -
      \frac{7}{8}\sin^3\alpha \right)\ ,
  \end{align}
\end{subequations}
which then applies also to the first-order energy (\ref{eq:IV.40})
\begin{equation}
  \label{eq:IV.44}
  \tEnT\{1\}\Rightarrow E_1(\alpha) =
  -\frac{e^2}{4\aB}\frac{(1+V_1\left(\alpha)\right)^2}{1+T_1(\alpha)}
  \doteqdot-\frac{e^2}{4\aB}\cdot S_1(\alpha)\ .
\end{equation}
Thus the energy eigenvalue problem (of first oder,~$\NN=1$) is reduced to the problem of
determining the stationary points of the \emph{spectral function}~$S_1(\alpha)$
without any constraint, see fig.~1.

Evidently, the first-order spectral function~$S_1(\alpha)$ (\ref{eq:IV.44}) has
period~$\pi$; and in this basic interval~$(0\le\alpha\le\pi)$ there exist four stationary
points, namely two maxima~(1a,1b) and two minima~(1c,1d). The corresponding
energies~$\tEnT\{1\}$ (\ref{eq:IV.44}) are the following

\parbox[c]{4cm}{\textbf{rel.\ maxima:}}
\parbox[c]{9cm}{
\begin{subequations}
  \begin{align}
    \label{eq:IV.45a}
    E_{1\mathrm{a}} &= - \frac{e^2}{4\aB}\cdot 1 \simeq -6,8029\ldots\ [eV]\\*
    \label{eq:IV.45b}
    E_{1\mathrm{b}} &= - \frac{e^2}{4\aB}\cdot 0,086717499\ldots\simeq -0,590\ [eV]
  \end{align}
\end{subequations}
}
- - - - - - - - - - - - - - - - - - - - - - - - - - - - - - - - - - - - - - - - - - - - - - - 

\parbox[c]{4cm}{\textbf{rel.\ minima:}}
\parbox[c]{9cm}{
\begin{subequations}
  \begin{align}
    \label{eq:IV.46a}
    E_{1\mathrm{c}} &= - \frac{e^2}{4\aB}\cdot 1,033319474 \ldots \simeq -7,030\ [eV]\\*
    \label{eq:IV.46b}
    E_{1\mathrm{d}} &= - \frac{e^2}{4\aB}\cdot 1,128194657\ldots\simeq -7,675\ [eV]\ .
  \end{align}
\end{subequations}
} 

Here, the first result~$E_{1a}$ (\ref{eq:IV.45a}) is clearly the most striking one,
because it agrees exactly with both the zero-order approximation (\ref{eq:IV.36}) and the
conventional groundstate prediction (\ref{eq:I.3}), where the latter is commonly
considered to be exact within the framework of the generally accepted non-relativistic
quantum mechanics. However, the occurrence of two minima and two maxima in the first-order
spectrum demands an explanation: From the physical point of view there should occur
for~$\NN=1$ just one maximum and one minimum corresponding to the groundstate and the first
excited state, as is actually the case with the conventional approach (see
Appendix). 

But evidently the present approximation procedure generates stationary points which
are \emph{spurious} in the sense that they are not present (not even approximately) in the
next-higher order of approximation. This viewpoint receives support from the subsequent
consideration of the second-order spectrum~$(\NN=2)$ which reproduces the energies
$E_{\mathrm{1b}}\ldots E_{\mathrm{1d}}$ (\ref{eq:IV.45b})-(\ref{eq:IV.46b}) but yields a
complete falsification of the maximum~$E_{\mathrm{1a}}$ (\ref{eq:IV.45a}). Here it should
be a matter of course that the increasing orders of approximation~$(\NN\to \NN+1\to
\NN+2\ldots)$ must be required to produce smaller and smaller corrections to the lower-order
predictions. This natural requirement is satisfied for the relative
minima~$E_{\mathrm{1c}}$ and~$E_{\mathrm{1d}}$ (\ref{eq:IV.46a})-(\ref{eq:IV.46b}) which
may be conceived as first-order successors of the single relative minimum (\ref{eq:IV.36})
of zero order; and thus the numerical coincidence of the first-order
\emph{maximum}~$E_{\mathrm{1a}}$ (\ref{eq:IV.45a}) with the zero-order \emph{minimum}
(\ref{eq:IV.36}) must appear as an incident. As a consequence for the present
approximation procedure, one is naturally led to a \emph{selection criterion}, namely in
the sense that for any approximation order~$\NN$ those stationary points have to be rejected
as artefacts which are not (even approximately) reproduced in the next
order~$\NN+1$. For~$\NN\to\infty$, the stationary points of infinite order would then
represent the true RST energy spectrum of non-relativistic positronium (in the spherically
symmetric approximation). This situation is in contrast to the conventional treatment (see
Appendix), where in any order~$\NN$ there arise just~$\NN+1$ stationary points which do
reappear in all higher orders~$(\NN\to\NN+1\to\NN+2\ldots)$ and therefore do represent
exact solutions of the conventional Schr\"odinger equation (\ref{eq:I.5}).

\clearpage

\begin{figure}
\begin{center}
\epsfig{file=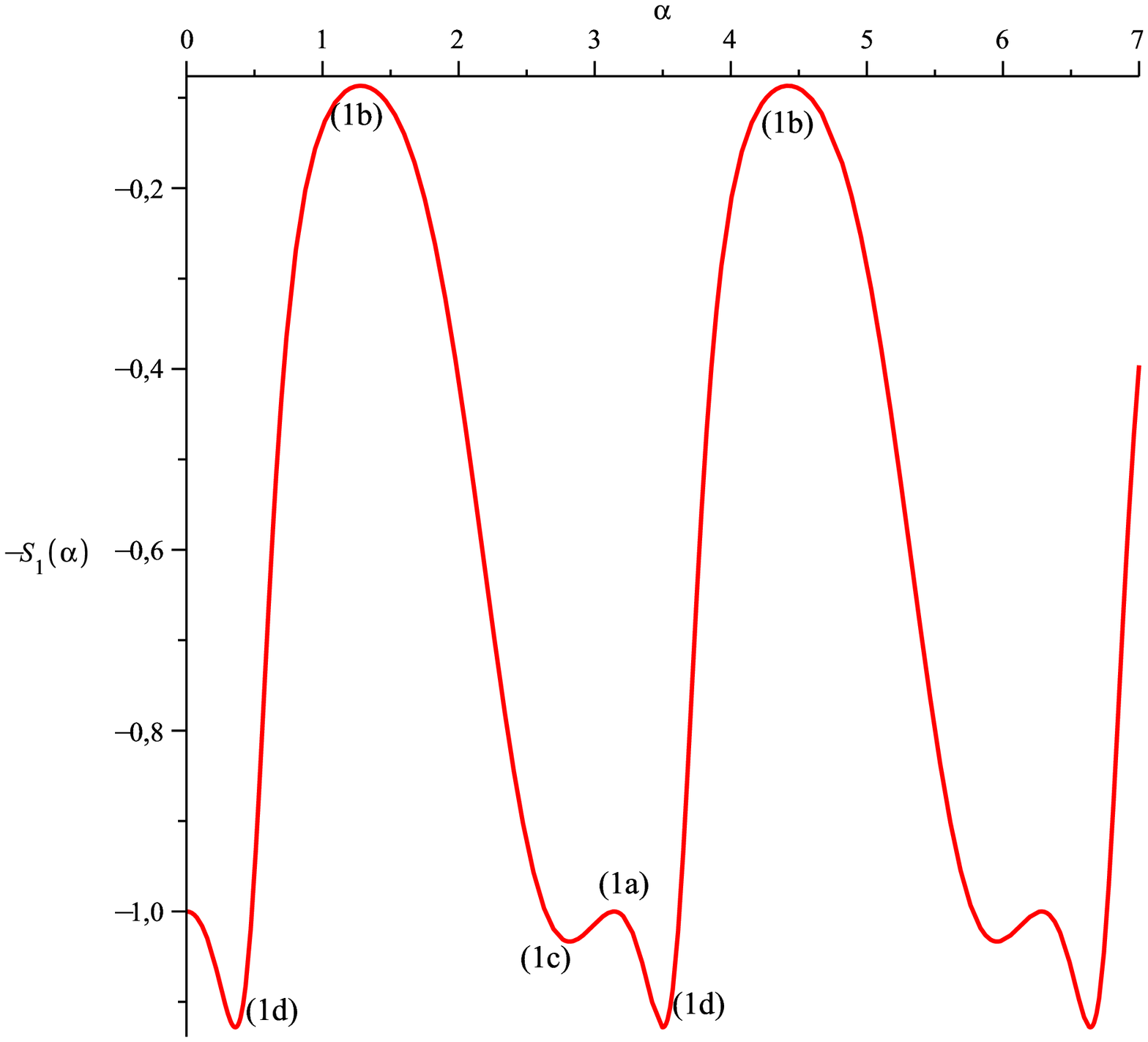,angle=0, width=13cm}
\end{center}
\end{figure}
\noindent
\textbf{Fig.\ 1:}\hspace{0.5cm} \emph{\large\bf First-Order Spectrum ($\NN=1$)}
\indent

Along the ``polynomial valley'' (\ref{eq:IV.41}), which is topologically equivalent to the
circle~$S^1\ (0\le\alpha\le 2\pi)$, there occur four different stationary points
(\ref{eq:IV.45a})-(\ref{eq:IV.46b}) of the energy functional~$\tEnT\{1\}$
(\ref{eq:IV.44}), namely two minima and two maxima. The maximum (1a) must be
rejected according to the selection criterion because this does not survive (as stationary
point) the transition to the next order~$(\NN=2)$ of the spectrum. The two minima (1c) and
(1d) represent first-order approximations of the true groundstate which is attainable only
in an asymptotic sense~$(\NN\to\infty)$.
\clearpage

\begin{center}
  \emph{\textbf{3.\ Second-Order Spectrum~$(\NN=2)$}}
\end{center}

For the second order~$(\NN=2)$ the general energy functional (\ref{eq:IV.27}) becomes
specialized to
\begin{equation}
  \label{eq:IV.47}
  \tEnT\{2\} = -\frac{e^2}{4\aB}\cdot\frac{\left(1+V_2\right)^2}{1+T_2}\ ,
\end{equation}
with the general kinetic function (\ref{eq:IV.23}) emerging here as
\begin{equation}
  \label{eq:IV.48}
  T_2(p_0,p_1,p_2) = -4p_1\left(p_0+p_1+10p_2\right)-16p_2\left(p_0+6p_2\right)\ ,
\end{equation}
and analogously the potential function (\ref{eq:IV.20}) is found as
\begin{equation}
  \label{eq:IV.49}
  \begin{split}
    V_2(p_0,p_1,p_2) &=
    p_0p_1\left(-2+9\cdot p_1^2+2\cdot p_0p_1+18\cdot p_0p_2+126\cdot p_1p_2 
      +627\cdot p_2^2\right)\\*
    &+p_0p_2\left(-10+42\cdot p_0p_2+1095\cdot p_2^2 \right)\\*
    &+p_1p_2\left(-51+201\cdot p_1^2+\frac{3057}{2}\cdot p_1p_2+\frac{10845}{2}\cdot p_2^2\right)\\*
    &-5\cdot p_1^2+\frac{21}{2}\cdot p_1^4-147\cdot p_2^2+\frac{15165}{2}\cdot p_2^4\ .
  \end{split}
\end{equation}
Furthermore, the general normalization condition (\ref{eq:IV.21}) adopts now the special
shape
\begin{equation}
  \label{eq:IV.50}
  p_0^2+4p_0p_1+6p_1^2+120 p_2^2+12p_0p_2+48p_1p_2=1
\end{equation}
which defines some compact submanifold  topologically equivalent to a two-sphere~$S^2$;
and therefore the following angular parameterization is feasible
\begin{subequations}
  \begin{align}
    \label{eq:IV.51a}
    p_0 &=
    \cos\alpha_1-\sqrt{2}\sin\alpha_1\cos\alpha_2+\sqrt{3}\sin\alpha_1\sin\alpha_2\\*
    \label{eq:IV.51b}
    p_1 &= \frac{\sqrt{2}}{2}\sin\alpha_1\cos\alpha_2 -
    \sqrt{3}\sin\alpha_1\sin\alpha_2\\*
    \label{eq:IV.51c}
    p_2 &= \frac{\sqrt{3}}{6}\sin\alpha_1\sin\alpha_2\\*
    &\hspace{3cm}\big(0\le\alpha_1\le\pi;\ 0\le\alpha_2\le 2\pi  \big)\ .\notag
  \end{align}
\end{subequations}
This converts again the kinetic function~$T_2(p_0,p_1,p_2)$ and the potential
function~$V_2(p_0,p_1,p_2)$ to the corresponding functions over the 2-sphere~$S^2$, i.e.
\begin{subequations}
  \begin{align}
    \label{eq:IV.52a}
    T_2(p_0,p_1,p_2)\Rightarrow T_2(\alpha_1,\alpha_2)\\*
    \label{eq:IV.52b}
    V_2(p_0,p_1,p_2)\Rightarrow V_2(\alpha_1,\alpha_2)\ .
  \end{align}
\end{subequations}
The stationary points of the spectral function~$S_2(\alpha_1,\alpha_2)$
\begin{equation}
  \label{eq:IV.53}
  S_2(\alpha_1,\alpha_2)\doteqdot\frac{\left(1+V_2(\alpha_1,\alpha_2)\right)}{1+T_2(\alpha_1,\alpha_2)}^2
\end{equation}
will then determine the energy~$\tEnT\{2\}$ (\ref{eq:IV.47}) again \emph{without} any constraint:
\begin{equation}
  \label{eq:IV.54}
  \tEnT\{2\}\Rightarrow E_2(\alpha_1,\alpha_2) = -\frac{e^2}{4\aB}\cdot
  S_2(\alpha_1,\alpha_2)\ .
\end{equation}
Clearly, this is now the second-order generalization of the first-order result
(\ref{eq:IV.44}).

The spectral function~$S_2(\alpha_1,\alpha_2)$ displays the occurrence of eight mirror
pairs of stationary points. This mirror arrangement is a consequence of the invariance of
the spectral function~$S_2(\alpha_1,\alpha_2)$ with respect to the inversion:
\begin{equation}
  \label{eq:IV.55}
  \{p_0,p_1,p_2\} \Rightarrow  \{-p_0,-p_1,-p_2 \}\ ,
\end{equation}
or expressed in terms of the spherical parameterization~$(\alpha_1,\alpha_2)$
\begin{subequations}
  \begin{align}
    \label{eq:IV.56a}
    \alpha_1&\Rightarrow \pi-\alpha_1\\*
    \label{eq:IV.56b}
    \alpha_2&\Rightarrow \pi+\alpha_2\ .
  \end{align}
\end{subequations}
Of course, this invariance of the energy functional~$ \tEnT$ with respect to the reflection
(\ref{eq:IV.55}) is nothing else than the fact that the wave function can be replaced by
its negative~$(\Psi\Rightarrow-\Psi)$ without changing the physics: both the mass
eigenvalue equations (\ref{eq:II.45a})-(\ref{eq:II.45b}) and the Poisson equations
(\ref{eq:II.43a})-(\ref{eq:II.43d}) are invariant with respect to the inversion
(\ref{eq:IV.55}). 

For the second order of approximation~$(\NN=2)$, our numerical program finds nine
stationary points (apart from reflection), where three of them do occur already in the
first order~$(\NN=1)$: these are the configurations~(1b) (\ref{eq:IV.45b}), (1c)
(\ref{eq:IV.46a}), and (1d) (\ref{eq:IV.46b}). This is a similar phenomenon as did occur
for proceeding from the zero-order approximation (\ref{eq:IV.36}) to the first-order
level~$E_{\mathrm{1a}}$ (\ref{eq:IV.45a}). However the latter configuration (1a)
unfortunately is the only first-order case which is not validated by the second-order
approach~$(\NN=2)$, despite the fact that it exactly agrees with the conventional
result! Instead, the six newly emerging configurations for~$\NN=2$ are the following:
\begin{subequations}
  \begin{align}
    \label{eq:IV.57a}
    E_{2\mathrm{e}} &= - \frac{e^2}{4\aB}\cdot 0,0259932\ldots\simeq -0,177\ldots\ [eV]\\*
    \label{eq:IV.57b}
    E_{2\mathrm{f}} &= - \frac{e^2}{4\aB}\cdot 0,0925612\ldots\simeq -0,629\ldots\ [eV]\\*
    \label{eq:IV.57c}
    E_{2\mathrm{g}} &= - \frac{e^2}{4\aB}\cdot 0,1130532\ldots\simeq -0,769\ldots\ [eV]\\*
   \label{eq:IV.57d}
    E_{2\mathrm{h}} &= - \frac{e^2}{4\aB}\cdot 1,0437688\ldots\simeq -7,100\ldots\ [eV]\\*
   \label{eq:IV.57e}
    E_{2\mathrm{i}} &= - \frac{e^2}{4\aB}\cdot 1,1286357\ldots\simeq -7,677\ldots\ [eV]\\*
   \label{eq:IV.57f}
    E_{2\mathrm{j}} &= - \frac{e^2}{4\aB}\cdot 1,1290162\ldots\simeq -7,680\ldots\ [eV]
  \end{align}
\end{subequations}
where the notation is chosen in such a way that~$E_{\mathrm{1b}}=E_{\mathrm{2b}},
E_{\mathrm{1c}}=E_{\mathrm{2c}}, E_{\mathrm{1d}}=E_{\mathrm{2d}}$. The comparison of this
second-order spectrum to the preceding first-order spectrum
(\ref{eq:IV.45a})-(\ref{eq:IV.46b}) demonstrates that the energy predictions are still
strongly changing when the approximation order~$\NN$ varies from~\mbox{$\NN=1$}
to~$\NN=2$. Thus one concludes that the order~$\NN$ has to be increased considerably in
order to get ``stable'' results (i.e.\ with negligible dependence upon~$\NN$). In contrast
to this, the conventional predictions are absolutely stable with respect to any chosen
order~$\NN$ (see Appendix).

The situation is more favorable with the minimal energies of both cases $\NN=1$ and
$\NN=2$. It is rather obvious that all five energy values\\ $(E_{\mathrm{1c}},
E_{\mathrm{1d}}, E_{\mathrm{2h}}, E_{\mathrm{2i}}, E_{\mathrm{2j}})$ are relatively close
to the ``exact'' value (\ref{eq:IV.36}). This suggests the conclusion that there exists a
true minimum at (roughly)~$-7,7\,[eV]$ so that the present~$\NN=1$ and~$\NN=2$ minima
yield a more or less good approximation to this lowest-energy state. Since the
conventional and experimental values are in the vicinity of (roughly)~$-6,803\ [eV]$,
there is a discrepancy of~$0,9\ [eV]$ with respect to the present spherically-symmetric
RST predictions. But this discrepancy is surely not due to the relativistic effects which
are of order~$\as^2\sim 10^{-3}\ldots 10^{-4}$. Thus it remains to be clarified whether
perhaps the present discrepancy is a consequence of the spherically symmetric
approximation (Sect.~III.3) or whether RST is intrinsically inaccurate. But here it is
interesting to note that a first estimate of the anisotropic effect~\cite{MaSo,BMS}
yielded a correction contribution of (roughly)~$1\ [eV]$ which is potentially just in the
right order of magnitude in order to let the ``exact'' non-relativistic RST prediction for
the positronium groundstate (practically) coincide with the conventional prediction
of~$-6,8\ [eV]$.

\begin{center}
  \emph{\textbf{4.\ Third-Order Approximation~$(\NN=3)$}}
\end{center}

On principle, the order of the spectrum could be increased arbitrarily provided the
computer capacity is large enough. Being satisfied with a certain lack of accuracy, our
computer program manages the case with~$\NN=3$ and finds the following energy levels:
\begin{subequations}
  \begin{align}
    \label{eq:IV.58a}
    E_{3\mathrm{a}} &= - \frac{e^2}{4\aB}\cdot 0,0112637\ldots\simeq -0,0766\ldots\ [eV]\\*
    \label{eq:IV.58b}
    E_{3\mathrm{b}} &= - \frac{e^2}{4\aB}\cdot 0,0259935\ldots\simeq -0,1768\ldots\ [eV]\\*
    \label{eq:IV.58c}
    E_{3\mathrm{c}} &= - \frac{e^2}{4\aB}\cdot 0,0283261\ldots\simeq -0,1926\ldots\ [eV]\\*
    \label{eq:IV.58d}
    E_{3\mathrm{d}} &= - \frac{e^2}{4\aB}\cdot 0,0382245\ldots\simeq -0,2600\ldots\ [eV]\\*
   \label{eq:IV.58e}
    E_{3\mathrm{e}} &= - \frac{e^2}{4\aB}\cdot 0,0925615\ldots\simeq -0,6294\ldots\ [eV]\\*
    \label{eq:IV.58f}
    E_{3\mathrm{f}} &= - \frac{e^2}{4\aB}\cdot 0,1130522\ldots\simeq -0,7690\ldots\ [eV]\\*
    \label{eq:IV.58g}
    E_{3\mathrm{g}} &= - \frac{e^2}{4\aB}\cdot 0,1159734\ldots\simeq -0,7889\ldots\ [eV]\\*
    \label{eq:IV.58h}
    E_{3\mathrm{h}} &= - \frac{e^2}{4\aB}\cdot 1,0437690\ldots\simeq -7,1006\ldots\ [eV]\\*
    \label{eq:IV.58i}
    E_{3\mathrm{i}} &= - \frac{e^2}{4\aB}\cdot 1,0484095\ldots\simeq -7,1322\ldots\ [eV]\\*
    \label{eq:IV.58j}
    E_{3\mathrm{j}} &= - \frac{e^2}{4\aB}\cdot 1,1286357\ldots\simeq -7,6779\ldots\ [eV]\\*
    \label{eq:IV.58k}
    E_{3\mathrm{k}} &= - \frac{e^2}{4\aB}\cdot 1,1288228\ldots\simeq -7,6792\ldots\ [eV]\\*
    \label{eq:IV.58l}
    E_{3\mathrm{l}} &= - \frac{e^2}{4\aB}\cdot 1,1290162\ldots\simeq -7,6805\ldots\ [eV]
  \end{align}
\end{subequations}
The general conclusion from such an arrangement is that all the energy levels
newly emerging in the order~$\NN$ will be present also in the order~$\NN+1$, while all the
levels of the \emph{preceding} order~$\NN-1$ have disappeared (check this also for the
passage from~$\NN=1$ (\ref{eq:IV.45a})-(\ref{eq:IV.46b}) to~$\NN=2$
(\ref{eq:IV.57a})-(\ref{eq:IV.57e})). For the present orders~$\NN=2$ and~$\NN=3$ the
common energy levels are the following:
\clearpage
\begin{subequations}
  \begin{align}
    \label{eq:IV.59a}
    E_{2\mathrm{e}} &\Leftrightarrow E_{3\mathrm{b}}\simeq -\frac{e^2}{4\aB}\cdot 0,0259935
    \ldots\simeq -0,176\ldots\ [eV]\\*
    \label{eq:IV.59b}
    E_{2\mathrm{f}} &\Leftrightarrow E_{3\mathrm{e}}\simeq -\frac{e^2}{4\aB}\cdot 0,092561
    \ldots\simeq -0,629\ldots\ [eV]\\*
    \label{eq:IV.59c}
    E_{2\mathrm{g}} &\Leftrightarrow E_{3\mathrm{f}}\simeq -\frac{e^2}{4\aB}\cdot 0,11305
    \ldots\simeq -0,769\ldots\ [eV]\\*
    \label{eq:IV.59d}
    E_{2\mathrm{h}} &\Leftrightarrow E_{3\mathrm{h}}\simeq -\frac{e^2}{4\aB}\cdot 1,04376
    \ldots\simeq -7,100\ldots\ [eV]\\*
    \label{eq:IV.59e}
    E_{2\mathrm{i}} &\Leftrightarrow E_{3\mathrm{j}}\simeq -\frac{e^2}{4\aB}\cdot 1,1286357
    \ldots\simeq -7,677\ldots\ [eV]\\*
    \label{eq:IV.59f}
    E_{2\mathrm{j}} &\Leftrightarrow E_{3\mathrm{l}}\simeq -\frac{e^2}{4\aB}\cdot 1,1290162
    \ldots\simeq -7,680\ldots\ [eV]
  \end{align}
\end{subequations}
For the relative arrangement of these third-order~$(\NN=3)$ RST results in comparison to
their conventional counterparts see fig.~2. This arrangement (logarithmic scale) seems to
support the hypothesis that the RST predictions will come even closer to their
conventional counterparts when the anisotropy of the RST interaction potential is taken
into account. It appears also somewhat amazing that the conventional levels for~$\np=1$
and~$\np=3$ should have no RST counterpart (at least up to the present order~$\NN=3$).
\clearpage
\begin{figure}
\begin{center}
\epsfig{file=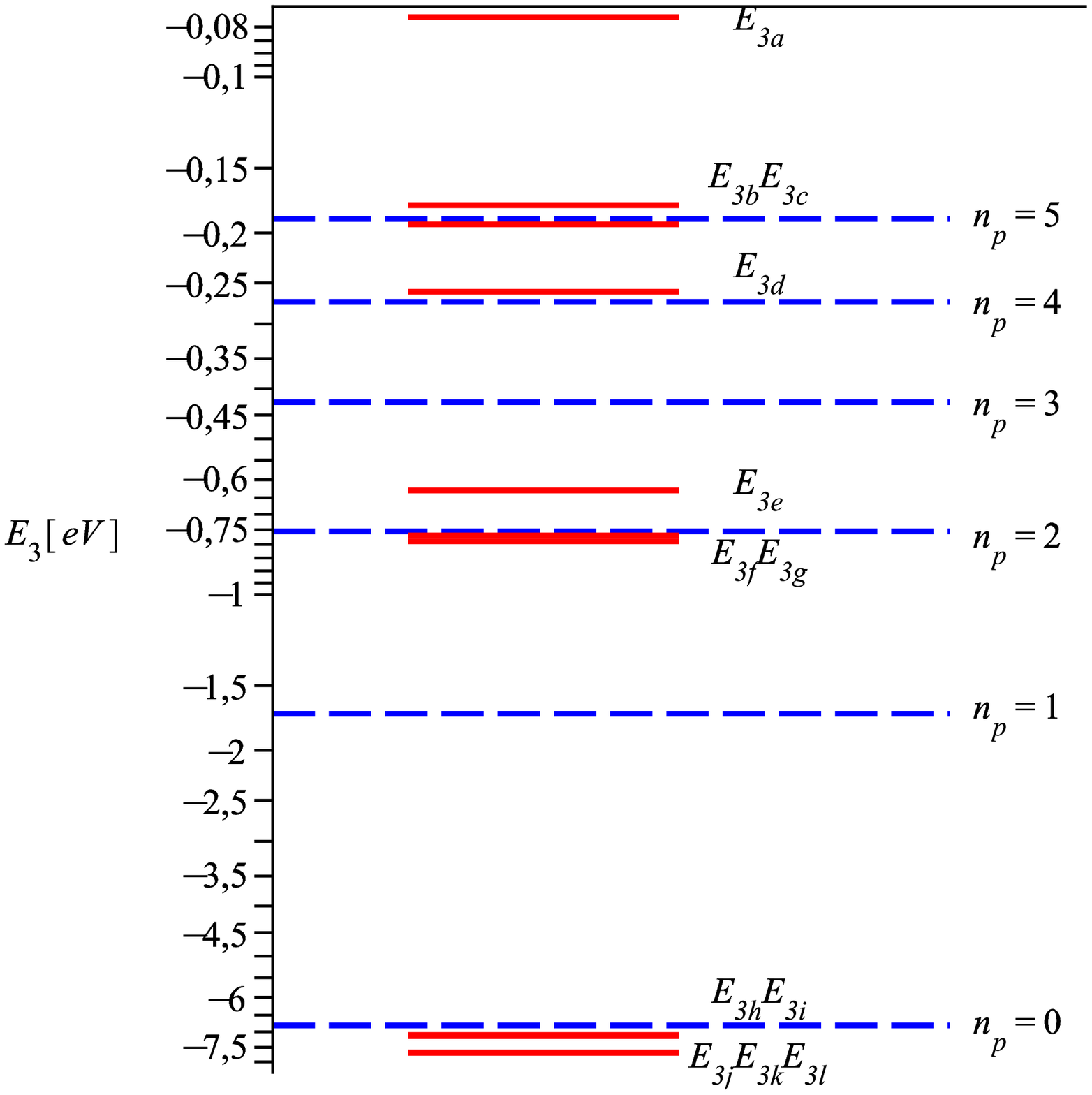,angle=0, width=13cm}
\end{center}
\end{figure}
\noindent
\textbf{Fig.\ 2:}\hspace{0.5cm} \emph{\large\bf RST Energy Levels of Third Order~($\NN=3$)}
\indent

The RST levels (\ref{eq:IV.58a})-(\ref{eq:IV.58l}) are relatively close (but not identical)
to the conventional levels~$(0\le\np\le 5)$ as must be expected for the spherically
symmetric approximation. The highest RST level~$E_{3\mathrm{a}}$ (\ref{eq:IV.58a}) is
inclusive because the order~$\NN$ must be increased considerably~$(\NN >> 3)$ in order to
describe such high energy levels. But the conventional groundstate~$(\np=0)$ is relatively
good approximated already in the present order~$\NN=3$.


\appendix
\section{Conventional Treatment}
\indent

The conventional treatment of the internal Coulomb force problem (\ref{eq:I.5}) yields
some further insight into the peculiarities of the variational approximation method of
Sect.~IV. Such a conventional approach relies upon the energy functional~$\Ec$
\begin{equation}
  \label{eq:A.1}
  \Ec = \frac{\hbar^2}{2m}\int d^3\vec{r}\left[\vec{\nabla}\phi_c(\vec{r}) \right]^2
  - e^2\int d^3\vec{r}\;\frac{\phi_c(\vec{r})^2}{r}
\end{equation}
whose variational equation is just the conventional Schr\"odinger equation (\ref{eq:I.5}),
provided the constraint of wave function normalization is respected
\begin{equation}
  \label{eq:A.2}
  \int d^3\vec{r}\;\phi_c(\vec{r})^2=1\ .
\end{equation}
Here it is convenient to recast the conventional wave function~$\phi_c(\vec{r})$ into dimensionless
form~$\phi(y)$:
\begin{gather}
  \label{eq:A.3}
  \phi_c(\vec{r}) = \frac{1}{\sqrt{4\pi\ab^3}}\cdot\phi(y)\\*
  \big(y\doteqdot \frac{r}{\ab} \big)\notag
\end{gather}
with the ``reduced'' Bohr radius~$\ab$
\begin{equation}
  \label{eq:A.4}
  \ab\doteqdot\frac{\hbar^2}{me^2}=2\aB
\end{equation}
being due to the reduced mass~$m=M/2$; and then the conventional functional (\ref{eq:A.1})
reappears in dimensionless form as
\begin{equation}
  \label{eq:A.5}
  \Ec = \frac{e^2}{\ab}\left[\frac{1}{2}\int dy\,y^2\left(\frac{d\phi(y)}{dy} \right)^2
  -\int dy\,y\cdot \phi(y)^2\right]\ ,
\end{equation}
where the original normalization condition (\ref{eq:A.2}) reads now in dimensionless form
\begin{equation}
  \label{eq:A.6}
  \int_0^\infty dy\,y^2\phi(y)^2 =1\ .
\end{equation}

Obviously, the present energy functional (\ref{eq:A.5}) with (\ref{eq:A.6}) is the
conventional analogue of the non-relativistic RST functional~$\tEnT$
(\ref{eq:III.32}). But here it is important to realize the crucial difference of both
approaches: whereas the conventional theory describes the mutual interaction by means of
the rigid Coulomb potential with no intrinsic dynamical freedom, RST adopts the interaction
potential~$\ppAn$ as a proper dynamical variable obeying its own field equations, even in
the non-relativistic approximation (see the Poisson equation (\ref{eq:III.34})).

But despite this structural difference of both approaches, the set of trial functions for
extremalizing either functional (\ref{eq:III.32}) or (\ref{eq:A.5}) may well be taken to
be the same, i.e.\ one may try the same functional form for the present conventional
treatment:
\begin{equation}
  \label{eq:A.7}
  \phi_\NN(y) = \sqrt{2\beta}P_\NN(y)e^{-\beta y} = \sqrt{2\beta}\left(\sum_{n=0}^\NN b_n y^n
  \right) e^{-\beta y}\ ,
\end{equation}
as was done for the preceding RST treatment, see (\ref{eq:III.35}) with
(\ref{eq:IV.1}). Defining also the modified polynomial coefficients~$p_m$ quite
analogously to the RST case (\ref{eq:IV.11})
\begin{equation}
  \label{eq:A.8}
  p_m\doteqdot\frac{b_m}{(2\beta)^{m+1}}\ ,
\end{equation}
the conventional normalization condition (\ref{eq:A.6}) then transcribes to these
coefficients as
\begin{equation}
  \label{eq:A.9}
  \sum_{m,n=0}^\NN p_m p_n (m+n+2)! = 1\ .
\end{equation}
Clearly, this defines again a compact submanifold of the configuration space spanned by
the hydrogen-like trial functions (\ref{eq:A.7}).

It should be obvious that the RST and the conventional approach share many common
features; and this becomes even more evident by considering now the action of the
conventional energy functional (\ref{eq:A.5}) upon the set of the hydrogen-like wave
functions. First, the kinetic part~($\cEk$, say) of the conventional energy~$\Ec$
(\ref{eq:A.5}) looks quite similar to the RST case (\ref{eq:IV.22}), namely
\begin{equation}
  \label{eq:A.10}
  \cEk\{\NN\} \doteqdot \frac{e^2}{2\ab}\int dy\,y^2\left(\frac{d\phi_\NN(y)}{dy}\right)^2
= \frac{e^2}{2\ab}\beta^2\left(1+\cTN\pN \right)
\end{equation}
where the conventional analogue~$\TN$ of the RST kinetic function~$\TN$
(\ref{eq:IV.23}) is given by
\begin{equation}
  \label{eq:A.11}
  \cTN\pN = -2\!\!\!\!\sum_{\stackrel{m,n=0} { (m+n)\ge 1} }^\NN p_m p_n (m+n)!\cdot\{m^2+n^2+n+m\}\ .
\end{equation}

But naturally, the conventional analogue~$(\cVN)$ of the RST potential function~$\VN$
(\ref{eq:IV.20}) will appear to be considerably simpler because, in the
conventional approach, both particles do interact instantaneously via the standard Coulomb
potential and this results in a purely quadratic potential function~$\cVN$. More concretely,
the potential energy contribution~$\Epot$ to~$\Ec$ (\ref{eq:A.5}) is given by
\begin{equation}
  \label{eq:A.12}
  \Epot\{\NN\} \doteqdot  -\frac{e^2}{\ab}\int dy\,y\cdot \phi_\NN(y)=-\frac{e^2}{\ab}\,\beta\left(
    1+\cVN\pN\right)\ ,
\end{equation}
where the conventional function~$\cVN$ is now found to be merely a \emph{quadratic} form of
the polynomial coefficients~$p_m$:
\begin{equation}
  \label{eq:A.13}
  \cVN\pN = -\sum_{\stackrel{m,n=0}{(m+n)\ge 1}}^\NN p_m p_n (m+n+1)!\cdot(m+n)\ .
\end{equation}

The comparison of this result to its RST analogue~$\VN$ (\ref{eq:IV.20}) says that the
dynamical character of the RST interaction is encoded in form of a quartic
contribution~$(\sim p_m p_n p_l p_q)$ to the potential function~$\VN$. Naturally, this
structural difference of both approaches will have its consequences for the
non-relativistic predictions of the positronium level system. The conventional
description may now be based upon the total energy function~$\Ec\{\NN\}$, i.e.
\begin{equation}
  \label{eq:A.14}
  \Ec\{\NN\} = \cEk\{\NN\} + \Epot\{\NN\} = \frac{e^2}{\ab}\left[\frac{1}{2}\beta^2
    \left(1+\cTN\right) - \beta\left(1+\cVN\right) \right]\ .
\end{equation}
This is the conventional analogue of the RST functional~$\tEnT\{\NN\}$ (\ref{eq:IV.24});
but unfortunately it is presently possible to analyze the value of this RST functional
(\ref{eq:IV.24}) only in the spherically symmetric approximation. In contrast to this, the
extremalization of the conventional functional~$\Ec\{\NN\}$ (\ref{eq:A.14}) is easily
feasible and corresponds to the ordinary Ritz variational principle~\cite{MaSo,BMS}.

\begin{center}
  \emph{\textbf{1.\ Zero-Order Approximation~$(\NN=0)$}}
\end{center}

The lowest-order polynomial~$P_0(y)$ (\ref{eq:A.7}) just consists of the lowest-order
coefficient~$b_0\ (=2\beta p_0)$ which, by means of the normalization condition
(\ref{eq:A.9}), is related to the variational parameter~$\beta$ through
\begin{equation}
  \label{eq:A.15}
  b_0=2\beta p_0=\sqrt{2}\beta\Rightarrow p_0=\frac{\sqrt{2}}{2}\ .
\end{equation}
Consequently, the normalized trial function (\ref{eq:A.7}) of zero order becomes
\begin{equation}
  \label{eq:A.16}
  \phi_0(y)=2\beta^{\frac{3}{2}} e^{-\beta y}\ ,
\end{equation}
while both the kinetic and potential functions do vanish
trivially~$(\T_0=\mathcal{V}_0=0)$. Thus the corresponding total energy~$\Ec\{0\}$
(\ref{eq:A.14}) adopts a very simple form:
\begin{equation}
  \label{eq:A.17}
  \Ec\{0\} = \frac{e^2}{\ab}\left(\frac{1}{2}\beta^2-\beta \right)
\end{equation}
whose minimal value is given by
\begin{equation}
  \label{eq:A.18}
  \frac{d\Ec\{0\}}{d\beta}=0\Rightarrow
  \Ec\big|_{\beta=1}=-\frac{e^2}{2\ab}=-\frac{e^2}{4\aB}\ .
\end{equation}
This just coincides with both the conventional Schr\"odinger result (\ref{eq:I.3}) and the
RST prediction~$\tEnT\{0\}$ of zero order (\ref{eq:IV.36}).

But clearly, one cannot be satisfied with such a zero-order success; and this forces one
to consider the next higher order.

\begin{center}
  \emph{\textbf{2.\ First-Order Spectrum~$(\NN=1)$ }}
\end{center}

The first-order wave function~$\phi_1(y)$ is deduced from the general expression
(\ref{eq:A.7}) as
\begin{equation}
  \label{eq:A.19}
  \phi_1(y)=\sqrt{2\beta}\left(b_0+b_1 y\right)e^{-\beta y} \ ,
\end{equation}
where the polynomial coefficients~$b_0,b_1$ (or~$p_0$ and~$p_1$, resp.) must satisfy the
normalization condition (\ref{eq:A.9}), i.e.\ for~$\NN=1$
\begin{equation}
  \label{eq:A.20}
  2p_0^2+12p_0p_1+24p_1^2=1\ .
\end{equation}
This is the conventional analogue of the former RST condition (\ref{eq:IV.41}) and
therefore can also be reparametrized by some angle~$\alpha\ (0\le\alpha\le2\pi)$:
\begin{subequations}
  \begin{align}
    \label{eq:A.21a}
    p_0 &= \sqrt{\frac{1}{2}}\cdot\cos\alpha-\sqrt{\frac{3}{2}}\cdot \sin\alpha\\*
    \label{eq:A.21b}
    p_1 &= \frac{\sin\alpha}{\sqrt{6}}\ ,
  \end{align}
\end{subequations}
cf.~the RST analogue hereof (\ref{eq:IV.42a})-(\ref{eq:IV.42b}).

As a consequence, both the kinetic and potential function~$\T_1(p_0,p_1)$
(\ref{eq:A.11}) and~$\mathcal{V}_1(p_0,p_1)$ (\ref{eq:A.13}), i.e.
\begin{subequations}
  \begin{align}
    \label{eq:A.22a}
    \T_1(p_0,p_1) &= -8p_1(p_0+2p_1)\\*
    \label{eq:A.22b}
    \mathcal{V}_1(p_0,p_1) &= -4p_1(p_0+3p_1)\ ,
  \end{align}
\end{subequations}
become now functions of that angle~$\alpha$:
\begin{subequations}
  \begin{align}
    \label{eq:A.23a}
    \mathcal{T}_1(p_0,p_1) &\Rightarrow \mathcal{T}_1(\alpha) = \frac{4}{3}(\sin^2\alpha
    -\sqrt{3}\sin\alpha\cos\alpha)\\*
    \label{eq:A.23b}
    \mathcal{V}_1(p_0,p_1) &\Rightarrow \mathcal{V}_1(\alpha) =
    -\frac{2}{\sqrt{3}}\sin\alpha\cos\alpha\ ,
  \end{align}
\end{subequations}
and the same does then apply also to the energy functional~$\Ec$ (\ref{eq:A.14})
\begin{equation}
  \label{eq:A.24}
  \Ec\{\NN\} \Rightarrow  \Ec\{1\}=\frac{e^2}{\ab}\left[\frac{1}{2}\beta^2\left(1+\mathcal{T}_1(\alpha)\right)  
  -\beta\left(1+\mathcal{V}_1(\alpha) \right) \right]\ .
\end{equation}
Here, the same procedure for eliminating the trial parameter~$\beta$ can be used as was
done for the RST case (\ref{eq:IV.25}), i.e.\ the equilibrium value for~$\beta$
\begin{equation}
  \label{eq:A.25}
  \beta = \frac{1+\mathcal{V}_1}{1+\mathcal{T}_1}
\end{equation}
is substituted back to~$\Ec\{1\}$ in order to yield the conventional energy
function~$\mathcal{E}_1(\alpha)$ as
\begin{equation}
  \label{eq:A.26}
  \mathcal{E}_1(\alpha) = -\frac{e^2}{4\aB}\cdot
  \frac{\left(1+\mathcal{V}_1(\alpha)\right)^2}{1+\mathcal{T}_1(\alpha)} 
  \doteqdot-\frac{e^2}{4\aB}\cdot \S_1(\alpha)\ .
\end{equation}

Of course, this is again the conventional counterpart of the former RST result
(\ref{eq:IV.44}); but the conventional spectral function~$\mathcal{S}_1(\alpha)$ (\ref{eq:A.26}), as
opposed to its RST counterpart~$S_1(\alpha)$ (fig.1), displays an important difference:
whereas the RST function~$S_1(\alpha)$ has two maxima (1a,1b) and two minima (1c,1d)
within the basic interval~$(0\le\alpha\le 2 \pi)$, the conventional
function~$\mathcal{S}_1(\alpha)$ (\ref{eq:A.26}) has only one minimum~(1A) and one
maximum~(1B), see fig.~A.1. The minimum occurs at~$\alpha_{\mathrm{1A}}=0\ (\mathrm{mod}\
\pi)$ and therefore the polynomial coefficients~$p_0,p_1$ (\ref{eq:A.21a})-(\ref{eq:A.21b})
become
\begin{subequations}
  \begin{align}
    \label{eq:A.27a}
    p_0\Big|_{\mathrm{1A}} &= \pm \sqrt{\frac{1}{2}}\\*
    \label{eq:A.27b}
    p_1\Big|_{\mathrm{1A}} &= 0\ ,
  \end{align}
\end{subequations}
which lets reappear the groundstate wave function~$\phi_0(y)$ (\ref{eq:A.16}) of the
zero-order approximation~$(\NN=0)$ also here on the first-order level~$\phi_1(y)$
(\ref{eq:A.19}). Clearly, with the minimum
value~$\mathcal{S}_1(\alpha)\Big|_{\mathrm{1A}}=1$ of the spectral
function~$\mathcal{S}_1(\alpha)$ one is again led back to the groundstate energy
(\ref{eq:A.18}) of the zero-order spectrum.

But a new element emerges on the first-order level~$(\NN=1)$ in form of the maximum~(1B),
see fig.~A1, which has the following polynomial coefficients~$p_0,p_1$
(\ref{eq:A.21a})-(\ref{eq:A.21b}):
\begin{subequations}
  \begin{align}
    \label{eq:A.28a}
    p_0\Big|_{\mathrm{1B}} &= -\frac{\sqrt{2}}{2}\\*
    \label{eq:A.28b}
    p_1\Big|_{\mathrm{1B}} &= \frac{\sqrt{2}}{4}\ .
  \end{align}
\end{subequations}
Furthermore, the associated exponential parameter~$\beta$ (\ref{eq:A.25}) becomes
\begin{equation}
  \label{eq:A.29}
  \beta\Big|_{\mathrm{1B}} = \frac{1}{2}
\end{equation}
so that the wave function~$\phi_1(y)$ (\ref{eq:A.19}) adopts the usual conventional
form for the standard Coulomb force problem
\begin{equation}
  \label{eq:A.30}
  \phi_1(y) \Rightarrow \phi_\mathrm{1B}(y) = -\frac{\sqrt{2}}{2} L_1(\frac{y}{2})e^{-\frac{1}{2}y}
\end{equation}
with~$L_1(\frac{y}{2})$ denoting the first Laguerre polynomial
\begin{equation}
  \label{eq:A.31}
  L_1(\frac{y}{2}) \doteqdot 1-\frac{y}{2}\ .
\end{equation}
Correspondingly, the energy~$\Ec\{1\}$ (\ref{eq:A.26}) is then found to coincide as
expected with the standard result (\ref{eq:I.3}) for the first excited state~$(\np=1)$ of the
ordinary Coulomb force problem:
\begin{equation}
  \label{eq:A.32}
  \Ec\Big|_{\mathrm{1B}} =
  -\frac{e^2}{4\aB}\cdot\frac{1}{2^2}=-\frac{e^2}{16\aB}\simeq-1,701\ldots\ [eV]\ .
\end{equation}
\clearpage
\begin{figure}
\begin{center}
\epsfig{file=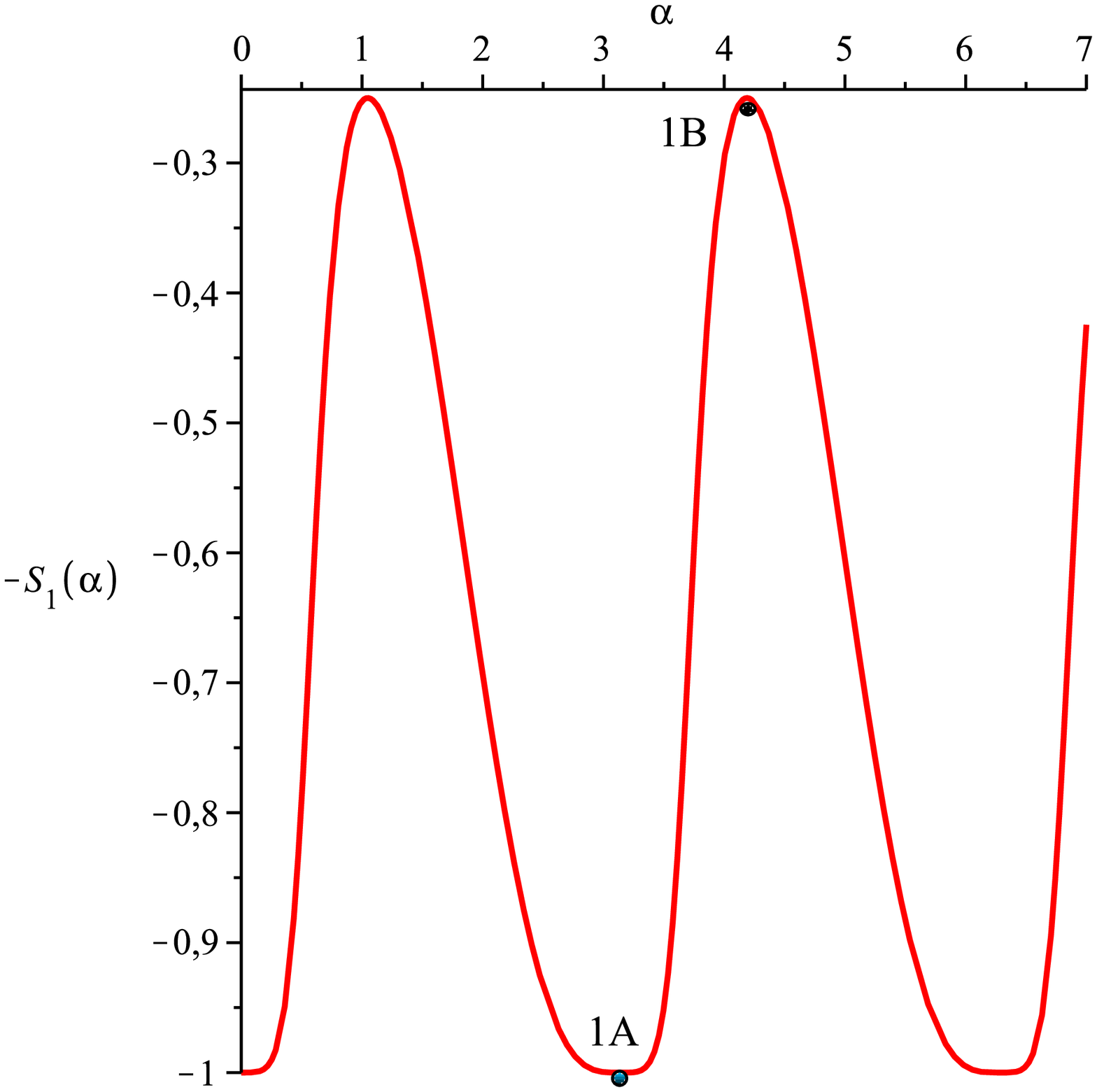,angle=0, width=13cm}
\end{center}
\end{figure}
\noindent
\textbf{Fig.\ A1:}\hspace{0.5cm} \emph{\large\bf Conventional First-Order Spectrum ($\NN=1$)}
\indent

The conventional spectral function~$\mathcal{S}_1(\alpha)$ (\ref{eq:A.26}) has one minimum
(1A) at~\mbox{$\alpha\big|_\mathrm{1A}=0\ (\mathrm{mod}\,\pi)$} and one maximum (1B)
at~$\alpha\big|_\mathrm{1B}=\frac{\pi}{3}\ (\mathrm{mod}\,\pi)$. This situation represents
the groundstate~($\leadsto$~1A) and first excited state~($\leadsto$~1B) of the
conventional theory, see equations (\ref{eq:A.18}) and (\ref{eq:A.32}).
\clearpage
\begin{center}
  \emph{\textbf{3.\ Second-Order Spectrum~$(\NN=2)$}}
\end{center}

It is both instructive and advantageous to reassure oneself that also for~$\NN=2$ the variational
process must take place over a compact manifold (i.e.\ the two-sphere~$S^2$). In order to
realize this clearly one starts with the normalization condition (\ref{eq:A.9}) for~$\NN=2$,
i.e.
\begin{equation}
  \label{eq:A.33}
  2p_0^2 + 12p_0p_1 + 24p_1^2 + 48p_0p_2 + 240p_1p_2 + 720p_2^2 = 1\ ,
\end{equation}
and satisfies this again by passing over to an angular parametrization; namely
\begin{subequations}
  \begin{align}
    \label{eq:A.34a}
    p_0 &= \sqrt{\frac{1}{2}}\cdot\cos\alpha_1 - \sqrt{\frac{3}{2}}\cdot\sin\alpha_1\cos\alpha_2
    +\sqrt{3}\sin\alpha_1\sin\alpha_2\\*
    p_1 &= \frac{\sin\alpha_1\cos\alpha_2}{\sqrt{6}} - \frac{2}{\sqrt{3}}\sin\alpha_1\sin\alpha_2\\*
    \label{eq:A.34c}
    p_2 &= \frac{\sin\alpha_1\sin\alpha_2}{4\sqrt{3}}\\*
    &\hspace{3cm}\Big(0\le\alpha_1\le\pi;0\le\alpha_2\le 2\pi \Big)\ .\notag
  \end{align}
\end{subequations}
Such a reparametrization converts the kinetic and potential
functions~$\mathcal{T}_2,\mathcal{V}_2$ (\ref{eq:A.11}) and (\ref{eq:A.13}) to the
corresponding angular functions
\begin{subequations}
  \begin{align}
    \label{eq:A.35a}
    \mathcal{T}_2(p_0,p_1,p_2) &= -\left(8p_0p_1 + 16p_1^2 + 48p_0p_2 + 192p_1p_2 + 576p_2^2 \right)
    \Rightarrow \T_2(\alpha_1,\alpha_2)\\*
    \label{eq:A.35b}
    \mathcal{V}_2(p_0,p_1,p_2) &= -\left(4p_0p_1 + 12p_1^2 + 24p_0p_2+144p_1p_2 + 480p_2^2 \right)
    \Rightarrow \mathcal{V}_2(\alpha_1,\alpha_2)\ .
  \end{align}
\end{subequations}

Correspondingly, the conventional second-order energy function~$\Ec\{2\}$ (\ref{eq:A.14})
can be recast again to the standard angular-dependent form by simply eliminating the exponential
parameter~$\beta$
\begin{equation}
  \label{eq:A.36} 
  \mathcal{E}_2(\alpha_1,\alpha_2) = -\frac{e^2}{4\aB}\frac{\left(1+\mathcal{V}_2(\alpha_1,\alpha_2)
    \right)^2}{1+\mathcal{T}_2(\alpha_1,\alpha_2)} \doteqdot - \frac{e^2}{4\aB}\cdot
  \mathcal{S}_2(\alpha_1,\alpha_2)\ .
\end{equation}
The stationary points over the two-sphere~$S^2$ can now be determined again without any
constraint and it should not come as a surprise that the second-order
spectral function~$\mathcal{S}_2$ (\ref{eq:A.36}) hast just three stationary points on the
sphere~$S^2$:
\begin{subequations}
  \begin{align}
    \label{eq:A.37a}
    \mathcal{S}_2\big|_{\mathrm{2A}} &= \mathcal{S}_1\big|_{\mathrm{1A}} = 1\\*
    \label{eq:A.37b}
    \mathcal{S}_2\big|_{\mathrm{2B}} &= \mathcal{S}_1\big|_{\mathrm{1B}} = \frac{1}{2^2}
    = \frac{1}{4}\\*
    \label{eq:A.37c}
    \mathcal{S}_2\big|_{\mathrm{2C}} &= \frac{1}{3^2} = \frac{1}{9}\ .
  \end{align}
\end{subequations}

Obviously the first two cases (\ref{eq:A.37a})-(\ref{eq:A.37b}) yield nothing else than
the groundstate (\ref{eq:A.18}) and the first excited state (\ref{eq:A.32}) which are
already present in the zero-order spectrum and in the first-order spectrum. Accordingly,
the stationary points coincide with the former ones (\ref{eq:A.15}),
(\ref{eq:A.27a})-(\ref{eq:A.27b}) and (\ref{eq:A.28a})-(\ref{eq:A.28b}), i.e.\ for the
groundstate~(A)
\begin{subequations}
  \begin{align}
    \label{eq:A.38a}
    p_0\big|_\mathrm{2A} &= p_0\big|_\mathrm{1A} = p_0\big|_0 = \pm \frac{\sqrt{2}}{2}\\*
    \label{eq:A.38b}
    p_1\big|_\mathrm{2A} &= p_1\big|_\mathrm{1A} = 0\ ,
  \end{align}
\end{subequations}
and similarly for the first excited state~(B)
\begin{subequations}
  \begin{align}
    \label{eq:A.39a}
    p_0\big|_\mathrm{2B} &= p_0\big|_\mathrm{1B} = \mp \frac{\sqrt{2}}{2}\\*
    \label{eq:A.39b}
    p_1\big|_\mathrm{2B} &= p_1\big|_\mathrm{1B} = \pm\frac{\sqrt{2}}{4}\\*
    \label{eq:A.39c}
    p_2\big|_{\mathrm{2B}} &= 0\ .
  \end{align}
\end{subequations}
Thus the second-order~$(\NN=2)$ wave functions for the two lowest-energy states are the same
as previously cf.~(\ref{eq:A.16}) and (\ref{eq:A.30})
\begin{subequations}
  \begin{align}
    \label{eq:A.40a}
    \phi_{\mathrm{2A}}(y) &= \phi_{\mathrm{1A}}(y) = \phi_0(y) = 2e^{-y}\\*
    \label{eq:A.40b}
    \phi_{\mathrm{2B}}(y) &= \phi_{\mathrm{1B}}(y) =
    -\frac{\sqrt{2}}{2}L_1\left(\frac{y}{2}\right)e^{-\frac{1}{2}y} \ .
  \end{align}
\end{subequations}

But a new element emerges in the second-order spectrum in form of the second excited
state~(2C) (\ref{eq:A.37c}). This state is due to the configuration
\begin{subequations}
  \begin{align}
    \label{eq:A.41a}
    p_0\big|_{\mathrm{2C}} &= \pm\frac{\sqrt{2}}{2},\quad
    p_1\big|_{\mathrm{2C}} = \mp\frac{\sqrt{2}}{2},\quad
    p_2\big|_{\mathrm{2C}} = \pm\frac{\sqrt{2}}{12}\\*
    \label{eq:A.41b}
    \beta\big|_{\mathrm{2C}} &= \frac{1}{3},\quad 
    \S_2\big|_{\mathrm{2C}} = \left(\frac{1}{3}\right)^2
  \end{align}
\end{subequations}
which yields the wave function
\begin{equation}
  \label{eq:A.42}
  \phi_{\mathrm{2C}}(y) = \pm\left(\frac{2}{3}\right)^{\frac{3}{2}}\left(1 -\frac{2}{3}y +
  \frac{2}{27}y^2\right)e^{-\frac{y}{3}}
\end{equation}
with the corresponding energy (\ref{eq:A.36})
\begin{equation}
  \label{eq:A.43}
  \Ec\big|_\mathrm{2C}=-\frac{e^2}{4\aB}\cdot\left(\frac{1}{3}\right)^2\simeq -0,76\ [eV]
\end{equation}
cf.\ the conventional spectrum (\ref{eq:I.3}) for~$\np=2$.

Summarizing, it should appear evident that the conventional spectrum of order~$\NN$ yields
just the series of~$\NN+1$ ``exact'' positronium levels (\ref{eq:I.3}) with~$0\le
n_p\le\NN$, together with their ``exact'' Schr\"odinger wave functions. Therefore the applied
variational method should be applicable also for calculating the \emph{non-relativistic}
positronium levels within the framework of RST. However a certain complication arises with
the latter approach because the RST levels become corrected by any higher
approximation~$(\NN\to \NN+1)$ whereas the conventional higher-order spectrum leaves
uncorrected the lower-order result. From the geometric viewpoint, this means that any
stationary point of the lower-dimensional configuration space remains a stationary point
after embedding into the next higher-dimensional configuration space~$(\NN\to
\NN+1\to\NN+2\to\ldots)$. Or conversely, the restriction of the conventional spectral
function~$\S_\NN(\alpha_1,\ldots\alpha_\NN)$ on the sphere~$S^\NN$ to the subsphere~$S^{\NN-1}$
does not generate new stationary points on the subsphere. Since this nice property is not
shared by the RST counterpart~$S_\NN(\alpha_1\ldots\alpha_\NN)$ it seems desirable to look for
a more rigorous mathematical characterization of those extraordinary functions being due
to both the conventional and RST type.



\end{document}